\documentclass[12pt,a4paper]{iopart}
\pdfoutput=1
\usepackage{iopams,setstack,mathrsfs,amsthm,cite,enumerate,float,graphicx}
\usepackage[utf8]{inputenc}
\usepackage[T1]{fontenc}
\usepackage{lmodern}
\usepackage[colorlinks,linkcolor=blue,citecolor=blue,urlcolor=blue]{hyperref}
\newcommand{\al}{\alpha}
\newcommand{\be}{\beta}
\newcommand{\de}{\delta}
\newcommand{\ep}{\epsilon}
\newcommand{\vep}{\varepsilon}
\newcommand{\ga}{\gamma}
\newcommand{\ka}{\kappa}
\newcommand{\la}{\lambda}
\newcommand{\om}{\omega}
\newcommand{\si}{\sigma}
\renewcommand{\th}{\theta}
\newcommand{\vp}{\varphi}

\newcommand{\ze}{\zeta}
\newcommand{\De}{\Delta}

\newcommand{\Om}{\Omega}
\newcommand{\bde}{\boldsymbol{\delta}}

\newcommand{\bbs}{\mathbf{s}}

\newcommand{\bT}{\mathbf{T}}

\newcommand{\bsi}{{\boldsymbol{\si}}}
\newcommand{\tb}{\widetilde{b}}

\newcommand{\hb}{\widehat{b}}

\newcommand{\NN}{{\mathbb N}}

\newcommand{\cB}{{\mathcal B}}

\newcommand{\cE}{{\mathcal E}}
\newcommand{\cF}{{\mathcal F}}
\newcommand{\cH}{{\mathcal H}}
\newcommand{\cI}{{\mathcal I}}

\newcommand{\cN}{{\mathcal N}}
\newcommand{\cP}{{\mathcal P}}

\newcommand{\cT}{{\mathcal T}}

\newcommand{\cW}{{\mathcal W}}
\newcommand{\cZ}{{\mathcal Z}}
\newcommand{\bb}{\overline b}

\newcommand{\pd}{\partial}

\newcommand{\ket}[1]{|#1\rangle}

\newcommand{\mss}{\kern 1pt}

\renewcommand{\le}{\leqslant}
\renewcommand{\ge}{\geqslant}
\newcommand{\tends}[1]{\bbuildrel{\hbox to 2em{\rightarrowfill}}_{#1}^{}}

\newcommand{\operatorname}[1]{\mathop{\rm #1}\nolimits}

\newcommand{\arcsinh}{\operatorname{arcsinh}}

\newcommand{\diff}{\mathrm{d}}

\newcommand{\ord}{\mathrm{o}}
\newcommand{\su}{\mathrm{su}}

\newcommand{\en}{\enspace}

\newcommand{\pdf}[2]{\frac{\partial #1}{\partial #2}}

\newcommand{\Int}[1]{\,\mathop{\!#1}\limits^{\lower1ex\hbox{$\scriptstyle\circ$}}{}}

\newcommand{\db}{b^\dagger}
\newcommand{\dc}{c^\dagger}

\newcommand{\df}{f^\dagger}
\newcommand\vru{\vrule width0pt height11pt depth6pt}

\theoremstyle{remark}

\let\tfrac\case
\let\eqref\eref

\newcommand{\dfrac}{\displaystyle\frac}
\eqnobysec
\def\clap#1{\hbox to 0pt{\hss#1\hss}}

\def\mathrlap{\mathpalette\mathrlapinternal}
\def\mathclap{\mathpalette\mathclapinternal}

\def\mathrlapinternal#1#2{%
           \rlap{$\mathsurround=0pt#1{#2}$}}
\def\mathclapinternal#1#2{%
           \clap{$\mathsurround=0pt#1{#2}$}}

\begin{document}

\title[Supersymmetric $t$-$J$ models with long-range interactions]{Supersymmetric $t$-$J$ models
  with long-range interactions: thermodynamics and criticality}

\author{B.~Basu-Mallick$^1$, N.~Bondyopadhaya$^2$, J.A.~Carrasco$^3$, F.~Finkel$^3$,
  A.~Gonz\'alez-L\'opez$^3$}

\address{$^1$Theory Division, Saha Institute of Nuclear Physics, HBNI, 1/AF Bidhan Nagar, Kolkata
  700 064, India} \address{$^2$Integrated Science Education and Research Centre, Siksha-Bhavana,
  Visva-Bharati, Santiniketan 731 235, India} \address{$^3$Departamento de F\'{\i}sica
  Te\'{o}rica, Universidad Complutense de Madrid, 28040 Madrid, Spain}

\eads{\mailto{bireswar.basumallick@saha.ac.in}, \mailto{nilanjan.iserc@visva-bharati.ac.in},
  \mailto{joseacar@ucm.es}, \mailto{ffinkel@ucm.es}, \mailto{artemio@ucm.es}}
\date{\today}
\begin{abstract}
  We analyze the thermodynamics and the critical behavior of the supersymmetric $\su(m)$ $t$-$J$
  model with long-range interactions. Using the transfer matrix formalism, we obtain a closed-form
  expression for the free energy per site both for a finite number of sites and in the
  thermodynamic limit. Our approach, which is different from the usual ones based on the
  asymptotic Bethe ansatz and generalized exclusion statistics, can in fact be applied to a large
  class of models whose spectrum is described in terms of supersymmetric Young tableaux and their
  associated Haldane motifs. In the simplest and most interesting $\su(2)$ case, we identify the
  five ground state phases of the model and derive the complete low-temperature asymptotic series
  of the free energy per site, the magnetization and charge densities, and their susceptibilities.
  We verify the model's characteristic spin-charge separation at low temperatures, and show that
  it holds to all orders in the asymptotic expansion. Using the low-temperature asymptotic
  expansions of the free energy, we also analyze the critical behavior of the model in each of its
  ground state phases. While the standard $\su(1|2)$ phase is described by two independent CFTs
  with central charge $c=1$ in correspondence with the spin and charge sectors, we find that the
  low-energy behavior of the $\su(2)$ and $\su(1|1)$ phases is that of a single $c=1$ CFT. We show
  that the model exhibits an even richer behavior on the boundary between zero-temperature
  phases, where it can be non-critical but gapless, critical in the spin sector but not in the
  charge one, or critical with central charge $c=3/2$.
\end{abstract}

\maketitle

\section{Introduction}
\label{sec.intro}

The $\su(m)$ $t$-$J$ model is one of the most intensively studied lattice models of strongly
correlated fermions, due to its relevance for the theoretical understanding of high-temperature
superconductivity and as one of the simplest quantum systems exhibiting spin-charge
separation~\cite{Sc87,ZR88,EKS92,Sc92}. The sites of this model can be occupied by at most one
charged fermion with $m$ internal degrees of freedom, which can hop between contiguous lattice
sites and interacts with its nearest neighbors through spin exchange and charge repulsion. The
one-dimensional $t$-$J$ model is of particular interest, since it is supersymmetric and exactly
solvable through the nested Bethe ansatz when its two parameters are suitably
related~\cite{Sc87,Su75,BB90,KY90,Sa91,EK92}.

In a recent paper~\cite{BBCFG19} we have computed in closed form the partition function of the
supersymmetric~$\su(m)$ $t$-$J$ model with long-range interactions introduced by Kuramoto and
Yokoyama~\cite{KY91,Ku93}. The lattice sites of the latter model are equispaced on a circle, and
each fermion can now interact with any other and hop among any two sites. Moreover, both the
interaction strength and the hopping amplitude are inversely proportional to the square of the
chord distance between the corresponding sites. The supersymmetric character of the $\su(m)$
Kuramoto--Yokoyama (KY) model can be established by mapping it to a suitable modification of the
$\su(1|m)$ Haldane--Shastry spin chain~\cite{Ha93}. This connection can in fact be exploited to
fully determine the spectrum of the former model in terms of supersymmetric motifs and their
corresponding Young tableaux~\cite{BBH10,BBCFG19}.

The thermodynamics of the supersymmetric KY model has been actively investigated ever since its
introduction. In fact, in the original reference~\cite{KY91} the low-temperature asymptotic
behavior of the magnetic and charge susceptibilities was determined by means of the asymptotic
Bethe ansatz (see, e.g.,~\cite{Su04}). A few years later, the thermodynamics of the $\su(m)$ KY
model at arbitrary temperature in the $N\to\infty$ limit was studied by Kato and
Kuramoto~\cite{KK96} applying Polychronakos's freezing trick~\cite{Po93} to the $\su(1|m)$
supersymmetric spin Sutherland model~\cite{KK95prl,AK96}. This method, which is rather involved,
requires first establishing the equivalence of the latter model to a system of
non-interacting~$\su(1|m)$ particles and then modding out the contribution of the dynamical
degrees of freedom. Moreover, it essentially relies on specific properties of the HS chain such as
its equivalence to a model of free particles with generalized momenta obeying fractional
statistics. On a more practical level, the formula for the grand potential obtained by Kato and
Kuramoto depends on a function which must be determined by solving an implicit equation with an
appropriate choice of branch.

In this paper we propose a novel direct method for analyzing the thermodynamics of the
supersymmetric KY model, which can be applied to a wide range of models with (complete or broken)
Yangian symmetry. We shall show how a formula for the grand potential of these models, akin to
Kato and Kuramoto's for the long-range supersymmetric $t$-$J$ model, emerges in a transparent way
from their partition function without requiring that they be described by generalized
pseudo-momenta or fractional statistics. In the simplest and most interesting case $m=2$, the
corresponding implicit equation is quadratic and can therefore be explicitly solved, which leads
to a new closed-form expression for the grand potential of the spin $1/2$ KY model.

The starting point in our method is the explicit formula for the partition function of the KY
model with an arbitrary (finite) number of sites~$N$ obtained in Ref.~\cite{BBCFG19}, which can be
recast into that of a related inhomogeneous vertex model. This key observation makes it feasible
to apply the transfer matrix method in Refs.~\cite{EFG12,FGLR18} to derive a closed-form
expression for the grand potential of the~$\su(m)$ KY model in the thermodynamic limit in terms of
the largest eigenvalue (in modulus) of a site-dependent transfer matrix. The characteristic
equation of this matrix, when expressed in an appropriate variable, is precisely the implicit
equation deduced by Kato and Kuramoto (henceforth referred to as the KK equation). In fact, our
method can be applied to any model described by an effective inhomogeneous vertex model, whose
energy function is expressible in terms of a dispersion relation and the supersymmetric Young
tableaux associated to finite-dimensional representations of the Yangian acting on tame
modules~\cite{KKN97,NT98,FG15}. By varying the dispersion relation we can derive the
thermodynamics of a large class of (partially or totally) Yangian-invariant systems, which
includes not only the KY model (or, equivalently, the $\su(1|m)$ supersymmetric HS chain) but
other well-known lattice models like the Polychronakos--Frahm (PF)~\cite{Po93,Fr93} or the
Frahm--Inozemtsev (FI)~\cite{FI94,BFGR10} spin chains. The grand potential of all of these models
can again be expressed in terms of the largest eigenvalue of a suitable transfer matrix. In the
$\su(1|m)$ case, we explicitly show that the characteristic equation of this transfer matrix is
equivalent to a generalized KK equation for a system of one boson and $m$ fermions. This strongly
suggests that the models in this class can be reformulated as systems of ``free'' $\su(1|m)$
particles (holons and spinons) interacting via appropriate fractional statistics. So far, this
conjecture has only been proved by an ad hoc method for the $\su(1|2)$ case in Ref.~\cite{KK96}.

A further aim of this paper is to take advantage of the explicit formula for the grand potential
of the spin~$1/2$ supersymmetric KY model in order to analyze in detail the low-temperature
behavior of its main thermodynamic functions, going beyond the first-order calculations in
Refs.~\cite{KK95,KK96}. To this end, we first determine the zero-temperature limit of the
magnetization and charge densities for all values of the magnetic field $h$ and the charge
chemical potential~$\mu$. In this way we identify the model's five ground state phases
in~$(h,\mu)$ space, characterized by their content of holes and fermions of both species. We then
compute the asymptotic expansion of the grand potential to all orders in~$T$, which turns out to
be different in each of these ground state phases. From the asymptotic series of the grand
potential we derive analogous infinite asymptotic expansions for the main thermodynamic functions
of interest, namely the magnetization and charge densities and their respective susceptibilities.
Apart from recovering the lowest-order results of Ref.~\cite{KK95}, we show that the strong
spin-charge separation characteristic of the model under consideration is a non-perturbative
property, in the sense that it persists at all orders in the low-temperature asymptotic expansion
of both susceptibilities. We also use our low-temperature asymptotic expansions to briefly analyze
the critical behavior of the spin $1/2$ KY model for arbitrary values of the magnetic field and
the charge chemical potential. In the genuinely $\su(1|2)$ phase we confirm the well-known result
that the model is described by two independent $c=1$ conformal field theories (CFT), one for each
of the spin and charge sectors~\cite{Ka92b}. On the other hand, in the $\su(2)$ and $\su(1|1)$
phases we interestingly find that the model, while still critical, is instead described by a
single $c=1$ CFT. The situation is even more complex on the boundary between ground-state phases,
where the model can be non-critical but gapless, critical in the spin sector but not in the charge
one, or have fractional central charge $c=3/2$.

The paper is organized as follows. In Section~\ref{sec.F} we introduce the model and recall from
Ref.~\cite{BBCFG19} its precise equivalence to (a modification of) the $\su(1|m)$ supersymmetric
Haldane--Shastry chain. We then exploit this equivalence to obtain explicit formulas for the free
energy per site and the main thermodynamic functions in the thermodynamic limit. In
Section~\ref{sec.KK96} we discuss the derivation of the Kato--Kuramoto equations and their
generalizations by means of the transfer matrix formalism. In the remaining sections we focus on
the simplest and most interesting case, namely the spin $1/2$ supersymmetric KY model. More
precisely, in Section~\ref{sec.T0} we obtain exact expressions for the zero-temperature
magnetization and charge densities, and apply them to identify the different ground state phases
in terms of the magnetic field strength and the charge chemical potential. Section~\ref{sec.FE} is
devoted to the derivation of the complete asymptotic series of the free energy per site in each of
the ground state phases, which we then use to analyze in detail the model's critical behavior. In
Section~\ref{sec.LTE} we compute the corresponding series for the magnetization per site, the
charge density and their susceptibilities, and discuss the spin-charge separation characteristic
of the model under study. We present our conclusions and outline some future developments in
Section~\ref{sec.conc}. The paper ends with three appendices in which we deal with several
technical questions arising in the derivation of the asymptotic series in Section~\ref{sec.FE}.

\section{Free energy of the $\su(m)$ Kuramoto--Yokoyama model}\label{sec.F}

\subsection{The model}

As shown in our previous paper~\cite{BBCFG19}, the Hamiltonian of the supersymmetric $\su(m)$ KY
model can be written as\footnote{Here and in what follows, unless otherwise stated, sums and
  products over Latin indexes run over the set~$1,\dots,N$ while Greek indices range from~$1$
  to~$m$.}
\begin{equation}
  \fl
  H_0=\frac{t\pi^2}{N^2}\,\sum_{i<j}\sin^{-2}\bigl(\tfrac\pi N\,(i-j)\bigr)\cP
  \bigg[-\sum_{\si}(\dc_{i\si}c_{j\si}+\dc_{j\si}c_{i\si})
  +2\bT_i\cdot\bT_j-\left(1-\tfrac1m\right)n_in_j\bigg]\cP.
  \label{tJSUSY}
\end{equation}
In the latter equations~$\dc_{i\si}$ (respectively~$c_{i\si}$) denotes the operator creating
(resp.~destroying) a fermion of type~$\si\in\{1,\dots,m\}$ at site~$i$ and
$n_i=\sum_\si n_{i\si}$, where $n_{i\si}=\dc_{i\si}c_{i\si}$, is the total number of fermions at
site~$i$. The operator $\cP$ is the projector onto single-occupancy states, in which each site is
occupied by at most one fermion. Finally, $\bT_i\equiv(T_i^1,\dots,T_i^{m^2-1})$, where~$T_i^r$ is
the~$r$-th $\su(m)$ Hermitian generator in the fundamental representation acting on the $i$-th
site. More precisely,
\begin{equation}\label{Tir}
  T_i^r=\sum_{\si,\si'}T^r_{\si\si'}\dc_{i\si}c_{i\si'}\,,
\end{equation}
where the complex numbers~$T^r_{\si\si'}$ are the matrix elements of the~$r$-th (Hermitian)
generator of~$\su(m)$ in the fundamental representation with the normalization
\[
  \tr(T^rT^s)=\frac12\,\de_{rs}\,.
\]
Thus the first term between braces in Eq.~\eqref{tJSUSY} accounts for the hopping of fermions
between sites~$i$ and~$j$, while the last two terms respectively model the spin (exchange) and
charge interaction between the latter sites.

As shown in Ref.~\cite{BBCFG19}, the KY Hamiltonian~\eqref{tJSUSY} can be mapped to (a suitable
modification of) the~$\su(1|m)$ Haldane--Shastry spin chain Hamiltonian by identifying the holes
of the KY model with the bosons of the HS spin chain. Indeed, the Hilbert space of the latter
chain is~$\hat\cH=\otimes_{i=1}^N\hat\cH_i$, where~$\hat\cH_i$ is the linear span of the
one-particle states $\db_i\ket{\hat\Om}_i$, $\df_{i\si}\ket{\hat\Om}_i$ ($\si=1,\dots,m$),
$\db_i$, $\df_{i\si}$ are the operators creating respectively a boson and a fermion of type~$\si$
at the $i$-th site and~$\ket{\hat\Om}_i$ is the vacuum in~$\hat\cH_i$. Similarly, let
$\cH=\otimes_{i=1}^N\cH_i$ denote the Hilbert space of the original model~\eqref{tJSUSY},
where~$\cH_i$ is the space spanned by its vacuum~$\ket\Om_i$ and the one-particle
states~$\dc_{i\si}\ket{\Om}_i$. The unitary mapping~$\vp:\cH\to\hat\cH$ defined by
\begin{equation}\label{phi}
  \vp\ket\Om_i=\db_i\ket{\hat\Om}_i\,,\quad
  \vp\big(\dc_{i\si}\ket\Om_i\big)=\df_{i\si}\ket{\hat\Om}_i
\end{equation}
induces a natural way of associating to each linear operator $A:\cH\to\cH$ a corresponding linear
operator~$\hat A=\vp A\vp^{-1}=\vp A \vp^\dagger$ acting on $\hat\cH$. It is shown in
Ref.~\cite{BBCFG19} that under this correspondence the Hamiltonian~$H_0$ in Eq.~\eqref{tJSUSY} is
transformed into
\begin{equation}
  \hat H_0
  =\frac{t\pi^2}{N^2}\,\bigg\{\sum_{i<j}\sin^{-2}\bigl(\tfrac\pi N\,(i-j)\bigr)
  (1-P_{ij}^{(1|m)})-
    \frac13\,(N^2-1)\,\cF\bigg\}\,,
  \label{hatH}
\end{equation}
where~$\cF\equiv\sum_{i}\hat n_{i}$ is the total number of $\su(1|m)$ fermions
and~$P_{ij}^{(1|m)}$ denotes the~$\su(1|m)$ supersymmetric permutation operator. Recall that the
action of~$P_{ij}^{(1|m)}$ on the canonical basis of~$\hat\cH$ is given by
\[
  P_{ij}^{(1|m)}\ket{\cdots\si_i\cdots\si_j\cdots}=\ep(\bsi)\ket{\cdots\si_j\cdots\si_i\cdots}\,,
\]
where $\ep(\bsi)$ is $1$ (respectively $-1$) if $\si_i=\si_j=0$ (resp.~$\si_i,\si_j\ge1$), while
for $\si_i\si_j=0$ and $\si_i\ne\si_j$ it is equal to the number of fermionic spins $\si_k$
with~$i+1\le k\le j-1$. Thus the first term in~$\hat H_0$ coincides with the Hamiltonian of the
$\su(1|m)$ supersymmetric HS chain~\cite{Ha88,Sh88,Ka92,Ha93,HH94}, as we had anticipated.

\subsection{Free energy}

We shall next explain how to compute in closed form the grand potential of the~$\su(m)$ KY
model~\eqref{tJSUSY} by exploiting its equivalence with the $\su(1|m)$ HS spin chain
Hamiltonian~\eqref{hatH}. In the thermodynamic limit, this is equivalent to computing the free
energy of the Hamiltonian
\begin{equation}
  \label{HKYmu}
  H=H_0-\frac12\sum_{\si=1}^{m-1}h_\si(n^\si-n^m)-\mu_c\sum_\si n^\si\equiv H_0+H_1,
\end{equation}
where $n^\si\equiv\sum_in_{i\si}$ denotes the total number of fermions of type~$\si$. The last
term in~$H_1$ is the chemical potential of the fermions (or, equivalently, of the electric
charge), while the first one can be interpreted as arising from the interaction with an external
$\su(m)$ magnetic field with strengths $h_1,\dots,h_{m-1}$ along each (Hermitian) generator of the
standard Cartan subalgebra of $\su(m)$ (see Ref.~\cite{BBCFG19} for more details). In particular,
for $m=2$ the term $-(h_1/2)(n^1-n^2)$ equals $-h_1S^z$, where~$S^z$ is the $z$ component of the
total spin operator. The $\su(1|m)$ spin chain Hamiltonian~$\hat H$ equivalent to $H$ under the
mapping~\eqref{phi} is~$\hat H=\hat H_0+\hat H_1$, where
\[
  \hat H_1=-\frac12\sum_{\si=1}^{m-1}h_\si(\cN_\si-\cN_m)-\mu_c\cF
\]
and~$\cN_\si\equiv\hat n^\si$ is the total numbers of~$\su(1|m)$ fermions of type~$\si$. We can
thus write
\begin{equation}
  \label{hatH0H1}
  \hat H=J H_{\mathrm{HS}}^{(1|m)}-\frac12\sum_{\si=1}^{m-1}h_\si(\cN_\si-\cN_m)-
  (t_0+\mu_c)\cF\,,
\end{equation}
where
\[
  H_{\mathrm{HS}}^{(1|m)}=\frac12\sum_{i<j}\sin^{-2}\bigl(\tfrac\pi N\,(i-j)\bigr)
  (1-P_{ij}^{(1|m)})
\]
and
\begin{equation}\label{t0HS}
  J=\frac{2t\pi^2}{N^2}\,,\qquad t_0=\frac{t\pi^2}{3N^2}\,(N^2-1)\,.
\end{equation}
The Hamiltonian~$\hat H$ can be more concisely expressed as
\begin{equation}
  \label{hatHmu}
  \hat H=J H_{\mathrm{HS}}^{(1|m)}-\sum_\si\mu_\si\cN_\si,
\end{equation}
where $\mu_\si$ is the chemical potential of the fermion of type~$\si$, given by
\begin{eqnarray}\label{musi}
  \mu_\si&=\frac12\,h_\si+\mu_c+t_0\,,\quad 1\le\si\le m-1\,;\\
  \mu_m&=-\frac12\,\sum_{\si=1}^{m-1}h_\si+\mu_c+t_0\,.
         \label{musim}
\end{eqnarray}
As shown in Refs.~\cite{BBCFG19} and~\cite{FGLR18}, the spectrum of the~$\su(1|m)$ spin
chain~\eqref{hatHmu}, and hence of the equivalent Hamiltonian~\eqref{HKYmu}, can be generated from
the formula
\begin{equation}\label{specH}
  E(\bbs)=J\sum_{i=1}^{N-1}\de(s_i,s_{i+1})i(N-i)
  -\sum_{i}\mu_{s_i}\,,
\end{equation}
where~$\mu_0\equiv0$, $\bbs\in\{0,\dots,m\}^N$ and~$\de(s,s')$ is defined by
\begin{equation}\label{de}
  \de(s,s')=\cases{1,&$s>s' \en\text{or}\en s=s'>0$\\
    0,&$s<s'\en\text{or}\en s=s'=0$\,.
  }
\end{equation}
The vectors~$\bde(\bbs)$ with components~$\de(s_i,s_{i+1})$ $(1\le i\le N-1)$ in Eq.~\eqref{specH}
are $\su(1|m)$ motifs~\cite{Ha93,HB00,BBH10}. In fact, the first sum in Eq.~\eqref{specH} can be
interpreted as the energy of a one-dimensional vertex model with~$N+1$ vertices~$0,\dots,N$ joined
by~$N$ bonds with values~$s_1,\dots,s_N\in\{0,\dots,m\}$, the energy associated to the $i$th
vertex~being equal to~$\de(s_i,s_{i+1})i(N-i)$~\cite{BBH10}.

Equation~\eqref{specH} is the key ingredient in the exact computation of the free energy in the
thermodynamic limit through the (site-dependent) transfer matrix method developed in
Ref.~\cite{FGLR18}. Indeed, from Eq.~\eqref{specH} it follows that the partition function can be
expressed as
\begin{equation}\label{cZA}
  \cZ=\tr\bigl(A(x_0)\cdots A(x_{N-1})\bigr),
\end{equation}
where $x_k\equiv k/N$ and the $(m+1)\times(m+1)$ transfer matrix~$A(x)$ has matrix elements
\begin{equation}\label{Aalbe}
  A_{\al\be}(x)=q^{K\vep(x)\de(\al,\be)-\frac12(\mu_{\al}+\mu_{\be})}\,,\quad 0\le\al,\be\le m\,.
\end{equation}
In the latter equation we have defined
\begin{equation}\label{qvepK}
  q=\e^{-1/T}\,,\qquad \vep(x)=x(1-x)\,,\qquad K=2t\pi^2>0\,,
\end{equation}
and as above we have taken (without loss of generality) $\mu_0\equiv0$. It can then be shown that
in the thermodynamic limit~$N\to\infty$ Eq.~\eqref{cZA} yields the following closed-form
expression for the free energy per site of the Hamiltonian~\eqref{HKYmu}:
\begin{equation}
  \label{f}
  f=-2T\int_0^{1/2}\log\la_1(x)\,\diff x\,,
\end{equation}
where~$\la_1(x)$ is the largest eigenvalue in modulus of the matrix~$A(x)$ (simple and positive,
by the Perron--Frobenius theorem). In fact, in the next section we shall explain how the latter
formula leads to the expression for the grand potential derived by a more laborious method in
Ref.~\cite{KK96}.

From now on we shall restrict ourselves to the~$\su(2)$ case, for which the eigenvalue~$\la_1(x)$
can be computed in closed form. Indeed, in this case the matrix~$A(x)$ is given by
\[
  A(x)=
  \left(
  \begin{array}{ccc}
    1& q^{-\mu_1/2}& q^{-\mu_2/2}\cr
    q^{K\vep(x)-\mu_1/2}& q^{K\vep(x)-\mu_1}& q^{-(\mu_1+\mu_2)/2}\cr
    q^{K\vep(x)-\mu_2/2}& q^{K\vep(x)-(\mu_1+\mu_2)/2}& q^{K\vep(x)-\mu_2}
  \end{array}
  \right).
\]
By Eqs.~\eqref{t0HS} and~\eqref{musi}-\eqref{musim}, the chemical
potentials~$\mu_\si$ are given (in the thermodynamic limit) by
\[
  \mu_\si=
  (-1)^{\si+1}\,\frac h2+\mu\,,
\]
where $h\equiv h_1$ and $\mu\equiv\mu_c+K/6$\,. Taking these relations into account, the
Perron--Frobenius eigenvalue of the matrix~$A(x)$ reads
\[
  \la_1(x)=\e^{\be(\mu-K\vep(x))}\Big[b(x)+\sqrt{b^2(x)+\e^{K\be\vep(x)}-1}\,\Big],
\]
where $\be\equiv1/T$ and
\begin{equation}
  \label{bKY}
  b(x)=\frac12\,\e^{\be(K\vep(x)-\mu)}+\cosh\bigl(\tfrac{\be h}2\bigr).
\end{equation}
The previous equations yield the following \emph{exact} explicit formula for the free energy per
site of the~$\su(2)$ KY model in the presence of a magnetic field~$h$ and charge chemical
potential~$\mu_c$:
\begin{equation}\label{fKY}
  f = -\mu_c-2T\int_0^{1/2}\kern-9pt\log\Bigl[b(x)+\sqrt{b^2(x)+\e^{K\be\vep(x)}-1}\,\Bigr] \diff
  x\,.
\end{equation}
The magnetization (per site)~$m_s=\langle n^1-n^2\rangle/(2N)$ and the charge (fermion)
density~$n_c=\langle n^1+n^2\rangle/N$ (where $\langle\cdot\rangle$ denotes thermal average) are
easily computed in closed form differentiating the latter equation, namely
\begin{eqnarray}
  \label{ms}
  m_s&=-\frac{\pd f}{\pd h}=\sinh\bigl(\tfrac{\be h}2\bigr)\int_0^{1/2}\kern-9pt
       D(x)^{-1/2}\diff
       x,\\
  n_c&=-\frac{\pd
       f}{\pd\mu_c}=1-\int_0^{1/2}\kern-9pt D(x)^{-1/2}\e^{\be(K\vep(x)-\mu)}\diff x,
       \label{nf}
\end{eqnarray}
with
\[
  D(x)\equiv b^2(x)+\e^{K\be\vep(x)}-1.
\]
The corresponding susceptibilities~$\chi_s$ and~$\chi_m$ are then given by
\begin{eqnarray}
  \fl
  \chi_s&=\frac{\pd m_s}{\pd h}=\frac\be2\int_0^{1/2}\kern-6pt
          \e^{\be(K\vep(x)-\mu)}D(x)^{-3/2}\bigg[1
          +\tfrac12\,\sinh^2\bigl(\tfrac{\be h}2\bigr)\nonumber\\
  \fl
  &\hphantom{=\frac{\pd n_c}{\pd\mu_c}=\be\int_0^{1/2}\kern-6pt\e^{\be(K\vep(x)-\mu)}D(x)^{-3/2}\bigg[
          \sinh^2}
        +\cosh\bigl(\tfrac{\be h}2\bigr)\Big(\e^{\be\mu}+\tfrac14\,\e^{\be(K\vep(x)-\mu)}\Big)
          \bigg]\diff x,
          \label{chis}\\
  \fl
  \chi_c&=\frac{\pd n_c}{\pd\mu_c}=\be\int_0^{1/2}\kern-6pt\e^{\be(K\vep(x)-\mu)}D(x)^{-3/2}\bigg[
          \sinh^2\bigl(\tfrac{\be h}2\bigr)\nonumber\\
  \fl
  &\hphantom{=\frac{\pd n_c}{\pd\mu_c}=\be\int_0^{1/2}\kern-6pt\e^{\be(K\vep(x)-\mu)}D(x)^{-3/2}\bigg[
          \sinh^2}
          +\tfrac12\,\e^{\be(K\vep(x)-\mu)}\cosh\bigl(\tfrac{\be
          h}2\bigr)+\e^{K\be\vep(x)}\bigg]\diff x.
  \label{chic}
\end{eqnarray}
  Other thermodynamic functions, like the energy~$u$, entropy~$s=\be(u-f)$ and specific heat (per
  site) $c_V=(\pd u)/(\pd T)$, are easily computed from Eq.~\eqref{fKY}.
  For instance,
  \begin{equation}
    \label{u}
    \fl
    u =\frac{\pd}{\pd\be}(\be
    f)=-\mu_c-\int_0^{1/2}\kern-9ptD(x)^{-1/2}\Big[h\sinh\bigl(\tfrac{\be
      h}2\bigr)+(K\vep(x)-\mu)\e^{\be(K\vep(x)-\mu)}\big)\Big]\,\diff x\,.
  \end{equation}

\section{Derivation of a generalized Kato--Kuramoto equation through the transfer matrix
  formalism}
\label{sec.KK96}

As explained in the Introduction, Kato and Kuramoto~\cite{KK96} obtained an expression for the
grand potential per spin~$\om$ of the KY Hamiltonian~$H_0$ in Eq.~\eqref{tJSUSY} in terms of a
function implicitly determined by an algebraic equation that we have named the Kato--Kuramoto
equation. It should be noted that the deduction of this equation in Ref.~\cite{KK96} requires that
(in our notation) the dispersion relation~$\vep(x)$ of the model be given by
Eq.~\eqref{qvepK}\footnote{In point of fact, in Ref.~\cite{KK96} there is a more general
  derivation of the KK equation based on the equivalence of the KY model to a system of $g$-ons
  with an appropriately chosen statistical matrix, but this derivation is valid only for the
  $\su(1|2)$ case.}. In this Section we shall first of all show how the KK equation emerges in a
transparent way from Eq.~\eqref{f} for the free energy per spin of the $\su(m)$ KY model with the
general chemical potential term $H_ 1$ in~Eq.~\eqref{HKYmu}. More importantly, we shall outline
how this equation can be generalized to a large class of solvable lattice models with (complete or
partial) Yangian invariance, including the supersymmetric PF and FI spin chains.

To this end, let us denote by~$P_m(\la)=\det(\la-A(x))$ the characteristic polynomial of the
transfer matrix~$A(x)$ in Eq.~\eqref{Aalbe}, where we have suppressed the dependence of~$P_m$ on
$x$ for the sake of conciseness. As remarked in Ref.~\cite{FGLR18} $P_m(\la)$ is divisible
by~$\la$, so that
\[
  Q_m(\la)=\frac{P_m(\la)}\la
\]
is an $m$-th degree monic polynomial. We shall next show that $Q_m$ satisfies the recursion
relation
\begin{equation}\label{recrel}
  Q_m(\la)=\la Q_{m-1}(\la)-\eta\,a_m^2\prod_{\si=1}^{m-1}\Big(\la+(1-\eta)a_{\si}^2\Big)\,,
  \qquad m\ge 2\,,
\end{equation}
where we have set
\[
  \eta=\e^{-K\be\vep(x)}\,,\qquad a_\si=\e^{\be\mu_\si/2}\,,\quad 1\le\si\le m\,.
\]
To see this, note that we can write
\[
  \fl
  (-1)^{m+1}P_m(\la)=
  \left|
  \begin{array}{cccccc}
    1-\la& a_1& a_2& \cdots & a_{m-1}&a_m\\
    \eta a_1& \eta a_1^2-\la& a_1a_2& \cdots& a_1a_{m-1}&a_1a_m\\
    \vdots& \vdots& \vdots& &\vdots&\vdots\\
    \eta a_{m-1}& \eta a_{m-1}a_1& \eta a_{m-1}a_2& \cdots& \eta a_{m-1}^2-\la& a_{m-1}a_m\\
    \eta a_{m}& \eta a_{m}a_1& \eta a_{m}a_2& \cdots& \eta a_ma_{m-1}& \eta a_m^2-\la
  \end{array}
  \right|.
\]
Multiplying the first row of the latter determinant by $\eta a_m$ and subtracting it from the last
one, after a straightforward calculation we obtain
\[
  \fl
  Q_m(\la)=\la Q_{m-1}(\la)-\eta a_m^2 \left|
  \begin{array}{cccccc}
    a_1& a_2& \cdots & a_{m-1}&1\\
    \eta a_1^2-\la& a_1a_2& \cdots& a_1a_{m-1}&a_1\\
    \vdots& \vdots& &\vdots&\vdots\\
    \eta a_{m-1}a_1& \eta a_{m-1}a_2& \cdots& \eta a_{m-1}^2-\la& a_{m-1}
  \end{array}
  \right|.
\]
The determinant in the previous formula can be easily evaluated by subtracting the last column
multiplied by $a_i$ from the $i$-th column for $i=1,\dots,m-1$, which yields Eq.~\eqref{recrel}.
From the latter recursion relation and the initial condition
\[
  Q_1(\la)=\la-\eta a_1^2-1
\]
we readily obtain the following explicit formula for~$Q_m(\la)$:
\[
  Q_m(\la)=\la^m-\la^{m-1}-\eta\sum_{k=0}^{m-1}\la^ka^2_{m-k}\prod_{\si=1}^{m-k-1}\Big(\la+(1-\eta)a_\si^2\Big).
\]
This expression can be somewhat simplified with the help of the identities
\[
  \prod_{\si=1}^{p}\Big(\la+(1-\eta)a_\si^2\Big)
  =\sum_{l=0}^{p}\la^l(1-\eta)^{p-l}e_{p-l}\bigl(a_1^2,\dots,a_{p}^2\bigr)
\]
and
\[
  e_k(x_1,\dots,x_{k+q})=\sum_{l=0}^{q}x_{k+l}e_{k-1}(x_1,\dots,x_{k+l-1})\,,
\]
where
\[
  e_k(x_1,\dots,x_{r})\equiv\sum_{1\le\si_1<\cdots<\si_{k}\le r}\kern-2em x_{\si_1}\cdots\, x_{\si_k}\
\]
denotes the elementary symmetric polynomial of degree~$k$ in $r\ge k$ variables. Indeed, after a
lengthy but straightforward calculation we obtain
\begin{equation}\label{laeq}
  (1-\eta)Q_m(\la)=\la^m-(1-\eta)\la^{m-1}-\eta\prod_{\si=1}^m\Big(\la+a_{\si}^2(1-\eta)\Big).
\end{equation}
Performing the change of variable
\begin{equation}\label{Xdef}
  X=\frac{1-\eta}\la,\qquad \la\ne0\,,
\end{equation}
we thus arrive at the fundamental identity
\begin{equation}\label{fundid}
  \la^{-m}(1-\eta)Q_m(\la)=1-X-\eta\prod_{\si=1}^m\Big(1+a_\si^2X\Big)\,,\qquad\la\ne0\,.
\end{equation}

In view of the previous discussion, we next rewrite Eq.~\eqref{f} for the free energy per site of
the supersymmetric~$\su(m)$ KY model with a chemical potential term as
\begin{equation}
  \label{KYX}
  f=2T\int_0^{1/2}\log\bigg(\frac{X_1(x)}{1-\e^{-K\be\vep(x)}}\bigg)\,\diff x\,,
\end{equation}
where
\begin{equation*}%\label{X1la1}
X_1(x)\equiv\frac{1-\e^{-K\be\vep(x)}}{\la_1(x)}\,.
\end{equation*}
Since~$\la_1(x)$ is a nonzero root of the characteristic equation of the matrix~$A(x)$, from
Eqs.~\eqref{Xdef}-\eqref{fundid} and the definition of~$\eta$ we deduce that the function~$X_1(x)$
satisfies
\[
  1-X_1=\eta\prod_{\si=1}^m\Big(1+a_\si^2X_1\Big)\,,
\]
or equivalently
\begin{equation}\label{xprel}
  K\be\vep(x)=-\log\bigl(1-X_1(x)\bigr)+\sum_\si\log\bigl(1+\e^{\be\mu_\si}X_1(x)\bigr)\,.
\end{equation}
Note that, since~$\la_1$ does not vanish and~$\vep(0)=0$, we must have
\begin{equation}\label{X10}
  X_1(0)=0\,.
\end{equation}
Equations~\eqref{KYX}-\eqref{X10} are equivalent to the expression for the grand potential per
site $\om$ of the $\su(m)$ KY model in Ref.~\cite{KK96}. To see this in more detail, note that in
the thermodynamic limit the Hamiltonian~$\hat H_0$ in Eq.~\eqref{hatH} (with~$t=1$) is related to
the analogous Hamiltonian
\begin{equation}\label{HtJKK}
  \cH_{t\text{-}J}=-\frac{\pi^2}{N^2}\sum_{i<j}\sin^{-2}\bigl(\tfrac\pi N\,(i-j)\bigr)P_{ij}^{(1|m)}
\end{equation}
in Ref.~\cite{KK96} by
\[
  \cH_{t\text{-}J}=\hat H_0+\frac{\pi^2}3\,\cF-\frac{N\pi^2}6,
\]
where we have used the identity
\begin{equation}\label{sinsum}
  \sum_{i\ne j}\sin^{-2}\bigl(\pi(i-j)/N\bigr)=\frac N3\,(N^2-1)\,.
\end{equation}
Thus in the thermodynamic limit the grand potential of $\cH_{t\text{-}J}$ should be equal to the
free energy of the right-hand side of Eq.~\eqref{HtJKK} with the addition of a chemical potential
term. Since we have absorbed the term proportional to~$\cF$ in the definition of the fermion
chemical potentials $\mu_\si$ (cf.~Eqs.~\eqref{musi}-\eqref{musim}), we must then show that
\begin{equation}\label{omx}
  \om=f-\frac{\pi^2}6=2T\int_0^{1/2}\log\biggl(\frac{X_1(x)}{1-\e^{-K\be\vep(x)}}\biggr)\,\diff
  x-\frac{\pi^2}6\,.
\end{equation}
Performing the change of variable~$x=(\pi-p)/(2\pi)$ in the integral, under which
\[
  K\vep(x)=2\pi^2x(1-x)=\frac12\,(\pi^2-p^2)\equiv\vep_0(p)\,,
\]
we see that it suffices to show that
\begin{equation}\label{om}
  \om=\frac{T}\pi\int_0^\pi\log\biggl(\frac{\tilde X_1(p)}{1-\e^{-\be\vep_0(p)}}\biggr)\,\diff
  p-\frac{\pi^2}6\,,
\end{equation}
where by Eqs.~\eqref{xprel}-\eqref{X10} $\tilde X_1(p)\equiv X_1(x(p))$ satisfies
\begin{equation}\label{xprel2}
  \be\vep_0(p)=-\log\bigl(1-\tilde X_1(p)\bigr)+\sum_\si\log\bigl(1+\e^{\be\mu_\si}\tilde X_1(p)\bigr)
\end{equation}
and
\begin{equation}\label{tX1pi}
\tilde X_1(\pi)=X_1(0)=0\,.
\end{equation}
Our claim now follows from the fact that Eqs.~\eqref{om} and~\eqref{xprel2} are nothing but
Eqs.~(2.33) and (2.22) in Ref.~\cite{KK96} (taking into account that we have set, without loss of
generality, $\mu_0=0$), while Eq.~\eqref{tX1pi} is the condition that according to the latter
reference determines the appropriate branch~$\tilde X_1(p)$ of the algebraic function defined by
Eq.~\eqref{xprel2}.

A few remarks on the equivalence of Eq.~\eqref{f} to~\eqref{om}-\eqref{tX1pi}--- or, more
generally, \eqref{KYX}-\eqref{X10}--- are now in order. The approach of Ref.~\cite{KK96} is based
on the derivation of the thermodynamics of the~$\su(1|m)$ spin Sutherland model in the
$N\to\infty$ limit, which yields the thermodynamics of the $\su(1|m)$ HS chain in the strong
coupling limit through Polychronakos's freezing trick. An essential ingredient in this approach is
the equivalence of the~$\su(1|m)$ spin Sutherland model to a system of non-interacting $\su(1|m)$
particles whose spectrum can be effectively described in terms of generalized momenta obeying an
appropriate exclusion statistics (see also~\cite{KK09}). From this description follows an integral
relation satisfied by the one-particle energy~$\tilde\vep(p)$ (defined
by~$\tilde X_1(p)=\e^{-\be\tilde\vep(p)}$), which in turn yields Eq.~\eqref{xprel2}.
Equation~\eqref{om} is then derived through a fairly elaborate argument, by first expressing the
grand potential of the spin Sutherland model in terms of the one-particle energy and then
subtracting the phonon contribution. By contrast, in our approach Eq.~\eqref{f} follows in a
straightforward way from Eqs.~\eqref{specH}-\eqref{de} applying the transfer matrix method, as
summarized in the previous section (see Ref.~\cite{FGLR18} for full details). That the spectrum of
the $\su(1|m)$ HS chain is completely described by equations analogous to~\eqref{specH}-\eqref{de}
is in fact a fundamental property of all $\su(n|m)$ spin chains of HS type, stemming directly from
the structure of their partition function~\cite{BBH10}. This description does not rely at all on
the properties of the associated spin Sutherland model, such as the existence of generalized
quasimomenta satisfying an appropriate exclusion statistics, and may thus apply even to other
types of Yangian-invariant models not necessarily derived from a dynamical spin model. In view of
the above argument, the free energy of all these models, including the three families of
$\su(n|m)$ spin chains of HS type, should also be described by equations analogous
to~\eqref{KYX}--\eqref{X10}. This is indeed remarkable, and it underscores the fact that the range
of applicability of the latter equations is much wider than could naively be expected from their
original derivation in Ref.~\cite{KK96}.

As a simple example of the last assertion, consider the $\su(n|m)$ Polychronakos--Frahm and
Frahm--Inozemtsev chains with a chemical potential term~$-\sum_{\al=1}^{m+n-1}\mu_\al\cN_\al$.
When $n=1$, the spectra of these models can be obtained replacing the Haldane--Shastry dispersion
relation~$i(N-i)$ in Eq.~\eqref{specH} respectively by $i$ and~$i(i+N\ga)$ (with
$N\ga>-1$)~\cite{EFG12,FGLR18}. It is then straightforward to show that if we set
\[
  K=\cases{NJ,\qquad&for the PF chain\\
  N^2J,&for the FI chain,
  }
\]
the partition function of these models is still given by
Eqs.~\eqref{cZA}-\eqref{Aalbe}, but with~$\vep(x)$ replaced by
\begin{equation}\label{epPFFI}
  \vep(x)= \cases{
    x\,,\qquad&for the PF chain\\
    x(x+\ga)\,, &for the FI chain
  }
\end{equation}
($\ga$ being now a nonnegative parameter). Consequently, in the thermodynamic limit their free
energy can be expressed in the form\footnote{The missing factor of $2$ and the different range of
  integration in Eq.~\eqref{fPFFI} compared to the analogous Eq.~\eqref{f} are due to the lack of
  symmetry about~$x=1/2$ of~$\vep(x)$ in Eq.~\eqref{epPFFI}.}
\begin{equation}\label{fPFFI}
  f=-T\int_0^1\log\la_1(x)\,\diff x\,,
\end{equation}
where~$\la_1(x)$ is the Perron--Frobenius eigenvalue of the transfer matrix~\eqref{Aalbe}
with~$\vep(x)$ as in Eq.~\eqref{epPFFI}. In the~$n=1$ case, from the latter equation we deduce
reasoning as above that the grand potential of the PF and FI chains can be written as
\begin{equation}\label{omPFFI}
  \om = T\int_0^1\log\biggl(\frac{X_1(x)}{1-\e^{-K\be\vep(x)}}\biggr)\,\diff
  x\,,
\end{equation}
where $X_1(x)$ satisfies the generalized KK equation~\eqref{xprel}-\eqref{X10} with~$\vep(x)$
given by~\eqref{epPFFI}. By the same token, in the general $\su(n|m)$ case with~$n>1$ we
conjecture that~\eqref{fPFFI} is equivalent to~\eqref{omPFFI}, where now $X_1(x)$ should satisfy
the generalized KK equation
\begin{equation}
  \label{genKK}
  \fl
  K\be\vep(x)=-\sum_{\al=0}^{n-1}\log\bigl(1-\e^{\be\mu_\al}X_1(x)\bigr)
  +\sum_{\al=n}^{m+n-1}\log\bigl(1+\e^{\be\mu_\al}X_1(x)\bigr)\quad
  (\text{with }\mu_0\equiv0)
\end{equation}
with the condition~$X_1(0)=0$. More generally, this should be true for \emph{any} model whose
spectrum be given by an equation of the form
\begin{equation}\label{Evertex}
  E(\bbs)=\frac{K}{N^\al}\sum_{i=1}^{N-1}\cE_N(i)\de(s_i,s_{i+1})
  -\sum_{i}\mu_{s_i}\,,
\end{equation}
together with the $\su(n|m)$ analogue of Eq.~\eqref{de}:
\begin{equation}\label{demn}
  \de(s,s')=\cases{1,&$s>s' \en\text{or}\en s=s'\ge n$\\
    0,&$s<s'\en\text{or}\en s=s'<n$\,,
  }
\end{equation}
provided that~$\lim_{N\to\infty}\cE_N(Nx)\equiv\vep(x)$ exists for all $x\in[0,1]$. In fact, an
equation akin to~\eqref{genKK} has been proposed in Ref.~\cite{KK09} for the supersymmetric
Sutherland model.

On a more technical level, our approach also helps clarify an important issue concerning the
definition of the function $\tilde X_1(p)$ through Eqs.~\eqref{xprel2}-\eqref{tX1pi}, or
equivalently~$X_1(x)$ through Eq.~\eqref{xprel} and the condition~$X_1(0)=0$. Indeed, from the
previous discussion it follows that the latter equation is equivalent to the algebraic equation
(with coefficients depending on the parameter~$x\in[0,1/2]$)
\begin{equation}\label{Qhat}
  \widehat Q(X)\equiv1-X-\e^{-K\be\vep(x)}\prod_{\si=1}^m\Big(1+\e^{\be\mu_\si}X\Big)=0
\end{equation}
for the variable~$X$. When $x=0$ it is clear that ~$X=0$ is a simple root of this equation, since
$\widehat Q$ vanishes at the origin and the coefficient of~$X$ in the latter polynomial is
$-1-\sum_{\si=1}^m\e^{\be\mu_\si}\ne0\,.$ By the implicit function theorem, the
condition~$X_1(0)=0$ uniquely defines a branch of the algebraic function~\eqref{Qhat} near~$x=0$.
However, it is not clear whether this is still the case ---i.e., whether there is no branch
crossing--- as $x$ increases. From a practical standpoint, the actual computation of $X_1(x)$
through Eqs.~\eqref{Qhat} and~\eqref{X10} at a point~$x>0$ is arduous at best, since it requires
following the appropriate branch of the algebraic function~\eqref{Qhat} all the way from $x=0$.
Both problems are solved by our approach, since it is now clear from Eq.~\eqref{Xdef} that
$X_1(x)$ is simply given by~$(1-\e^{-K\be\vep(x)})/\la_1(x)$, where $\la_1(x)$ is the eigenvalue
of the matrix $A(x)$ with the largest modulus (whose uniqueness for arbitrary $x$ is guaranteed by
the Perron--Frobenius theorem).

\section{Ground state phases}\label{sec.T0}

In this section we shall use Eqs.~\eqref{ms}-\eqref{nf} to derive exact expressions for the
zero-temperature magnetization and charge densities of the spin $1/2$ KY model for arbitrary
values of the magnetic field strength $h$ and charge chemical potential~$\mu$. In this way we
shall identify the model's five ground state phases, which in turn determine the form of the
low-temperature asymptotic series that we shall compute in the next section.

\begin{figure}[h]
  \includegraphics[width=.49\textwidth]{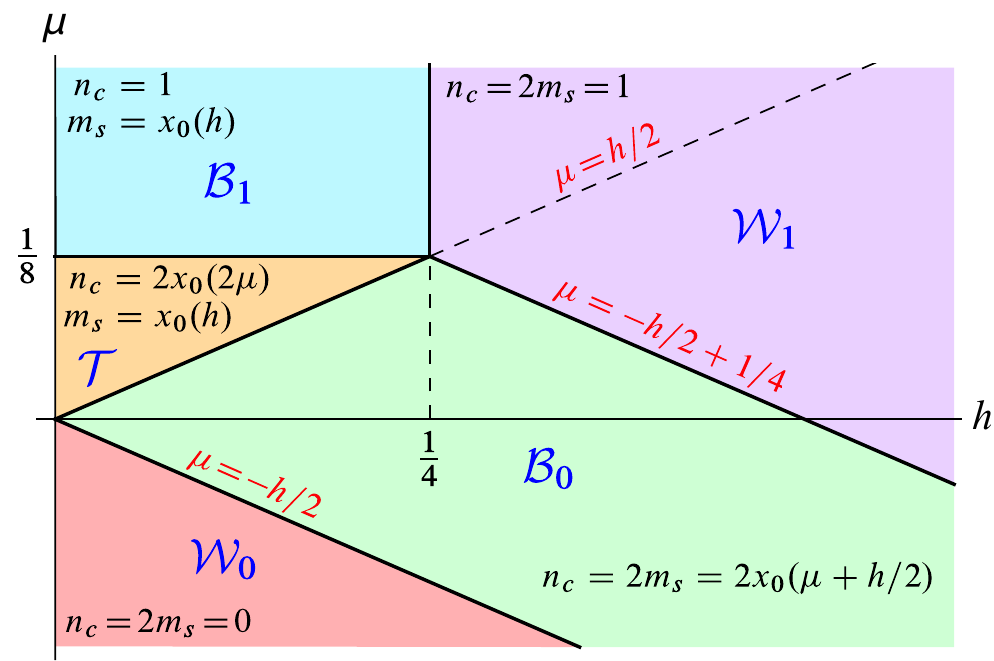}\hfill
  \raisebox{-4pt}{%
  \includegraphics[width=.49\textwidth]{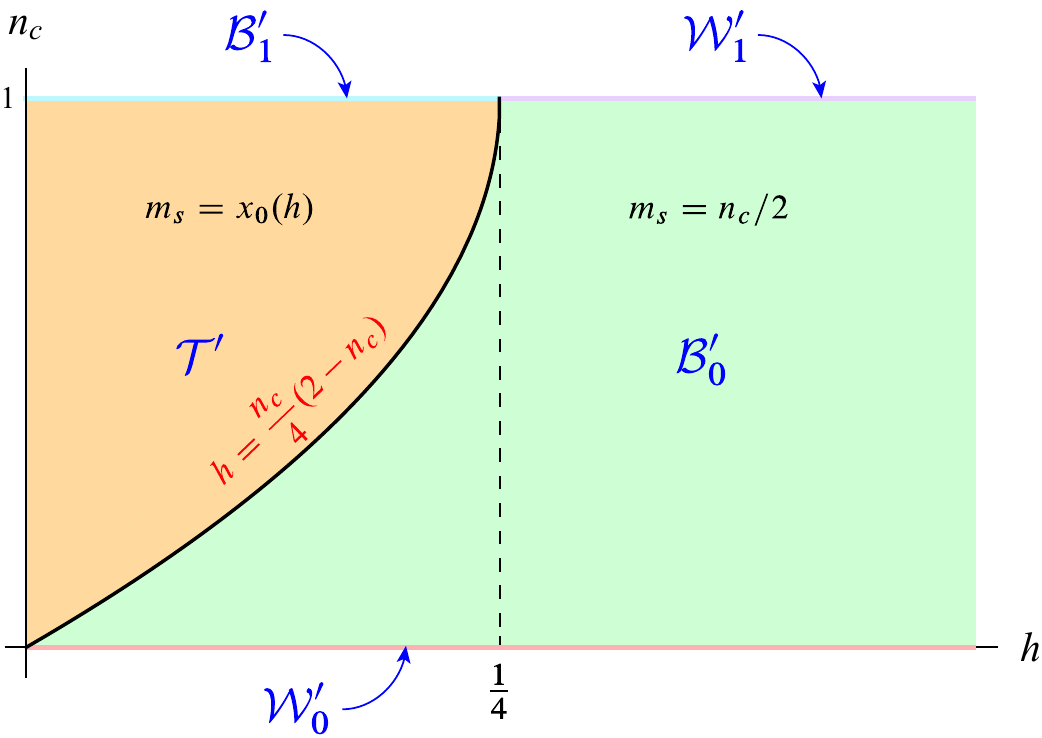}}
  \caption{Left: zero temperature magnetization and charge densities as functions in each of the
    regions $\cB_i$, $\cT$, $\cW_i$ defined in Table~\ref{table.regs}. Right: Images $\cB_i'$,
    $\cT'$, $\cW_i'$ of the regions $\cB_i$, $\cT$, $\cW_i$ under the mapping
    $(h,\mu)\mapsto(h,n_c)$.}
  \label{fig.hmuhnc}
\end{figure}

In order to simplify the calculations, in the rest of the paper we shall take~$K$ as the unit of
energy and hence of temperature (since~$k_{\mathrm B}=1$ from the outset). We can also suppose
without loss of generality that~$h\ge0$, since changing the sign of~$h$ is equivalent to
exchanging $n^1$ with $n^2$, or equivalently replacing~$(m_s,n_c)$ by~$(-m_s,n_c)$. The
magnetization per site~$m_s$ clearly vanishes for~$h=0$. On the other hand, for~$h>0$ and~$T\to0$
we can replace~$\sinh(\be h/2)$ and~$\cosh(\be h/2)$ by $\e^{\be h/2}/2$ up to an exponentially
small term~$\Or(\e^{-\be h/2})$. Hence at low temperatures we can write
\begin{equation}\label{msTsmall}
  m_s\simeq \int_0^{1/2}\kern-4pt\big[4\e^{-\be h}D(x)\big]^{-1/2}\diff x,
\end{equation}
with
\[
  4\e^{-\be h}D(x)\simeq1+4\e^{\be(\vep(x)-h)} +2\e^{\be(\vep(x)-\mu-\frac
    h2)}+\e^{2\be(\vep(x)-\mu-\frac h2)}\,,
\]
where we have dropped several exponentially small terms $\Or(\e^{-\be h})$. From the previous
equations it immediately follows that the zero-temperature magnetization is given by
\[
  m_s=\Big|\{x\in[0,\tfrac12]:\vep(x)<h,\ \vep(x)<\mu+\tfrac h2\}\Big|\,,
\]
where~$|A|$ denotes the measure of the set~$A$. Thus, at~$T=0$ we have
\begin{equation}\label{mshmu}
  m_s=
  \cases{
    x_0(h),& $(h,\mu)\in \cB_1\cup \cT$\\
    \frac12,& $(h,\mu)\in \cW_1$\\
    x_0\bigl(\mu+\tfrac h{2}\bigr),& $(h,\mu)\in \cB_0$\\
    0,&$(h,\mu)\in \cW_0$\,,
  }
\end{equation}
where
\begin{equation}\label{x0t}
  x_0(t)=\frac12\,\big(1-\sqrt{1-4t}\,\big)
\end{equation}
is the unique root of the equation~$\vep(x)=t$ in the interval~$[0,1/2]$, and the regions $\cB_i$,
$\cT$ and $\cW_i$ are defined in Table~\ref{table.regs} (cf.~Fig.~\ref{fig.hmuhnc} left). It is
also straightforward to check that~$m_s$ is continuous on the boundaries of the latter sets, and
hence everywhere.
\begin{table}
  \centering
  \begin{tabular}{|c|c|c|}\hline
    Region & Equation& Species content\\ \hline\hline
    $\cB_0$ & $\,0<\mu+\frac h2<\frac 14$,\quad $\mu<\frac h2\,$& Bosons and ``up'' fermions\vru\\ \hline
    $\cB_1$ & $h< \frac 14$,\quad $\mu>\frac 18$& Fermions\vru\\ \hline
    $\cT$ & $\frac h2<\mu<\frac 18$& Bosons and fermions\vru\\ \hline
    $\cW_0$ & $\mu<-\frac h2$& Bosons\vru\\ \hline
    $\cW_1$ & $h>\frac 14$,\quad $\mu>-\frac h2+\frac 14$& ``Up'' fermions\vru\\ \hline
  \end{tabular}
  \caption{Definitions of the regions $\cB_i$, $\cT$ and $\cW_i$ in the half-plane $h\ge0$
    (cf.~Fig.~\ref{fig.hmuhnc} left) and their species content.}\label{table.regs}
\end{table}

Similarly,
\[
  1-n_c=2\int_0^{1/2}[4\e^{-2\be(\vep(x)-\mu)}D(x)]^{-1/2}\diff x,
\]
where the term in brackets can be approximated at low temperatures by
\begin{eqnarray*}
  \fl
  4\e^{-2\be(\vep(x)-\mu)}&(\e^{\be\vep(x)}-1)
                            +\big(1+\e^{-\be(\vep(x)-\mu-\frac h2)}\big)^2\\
  \fl
  &=1+\e^{-2\be(\vep(x)-\mu-\frac h2)}+ 2\e^{-\be(\vep(x)-\mu-\frac
    h2)}+4\e^{-\be(\vep(x)-2\mu)}-4\e^{-2\be(\vep(x)-\mu)}
\end{eqnarray*}
Proceeding as before we obtain the following expression for the zero-temperature charge density:
\[
  \frac12\big(1-n_c\big)=\Big|\{x\in[0,\tfrac12]:\vep(x)>2\mu,\ \vep(x)>\mu+\tfrac h2\}\Big|,
\]
or equivalently
\[
  n_c=2\,\Big|\{x\in[0,\tfrac12]:\vep(x)<2\mu\ \text{or}\ \vep(x)<\mu+\tfrac h2\}\Big|.
\]
Thus the zero-temperature charge density is given by
\begin{equation}\label{nc0}
  n_c=
  \cases{
    1,& $(h,\mu)\in \cB_1\cup \cW_1$\\
    2x_0(2\mu),& $(h,\mu)\in \cT$\\
    2x_0(\mu+\tfrac h{2}\bigr),& $(h,\mu)\in \cB_0$\\
    0,& $(h,\mu)\in \cW_0$\,.
  }
\end{equation}
As before, it is easily verified that $n_c$ is continuous across the boundaries of the regions
$\cB_i$, $\cW_i$, $\cT$. It should also be noted that Eqs.~\eqref{mshmu}-\eqref{nc0} were derived
in Ref.~\cite{BBCFG19} by a more laborious method, based on determining the magnon content of the
ground state for arbitrary values of the parameters~$h$ and~$\mu$ using
Eqs.~\eqref{specH}-\eqref{de} for the energies of the equivalent vertex model.

As remarked in Ref.~\cite{BBCFG19}, the regions~$\cB_i$, $\cT$ and~$\cW_i$ defined above have a
clear interpretation as different zero-temperature phases of the model. Indeed, taking into
account that
\[
  m_s=\frac1{2N}\,\langle n^1-n^2\rangle\,,\qquad n_c=\frac1N\,\langle n^1+n^2\rangle\,,
\]
it is clear the latter regions are characterized by their different species content as listed in
Table~\ref{table.regs}. Thus in the bands~$\cB_0$ and $\cB_1$ the spin~$1/2$ KY model is
respectively equivalent (at zero temperature) to an~$\su(1|1)$ and an (antiferromagnetic) $\su(2)$
Haldane--Shastry chain, while in the wedges~$\cW_i$ it is trivial. In fact, the value of the
zero-temperature magnetization in the bands~$\cB_{0,1}$ coincides with the corresponding one for
the~$\su(1|1)$ and $\su(2)$ HS spin chains computed in Refs.~\cite{CFGRT16} and~\cite{EFG12}. On
the other hand, in the triangle~$\cT$ the model is genuinely of~$\su(1|2)$ type. We shall see in
the next sections that in each of the zero-temperature phases found above the thermodynamic
quantities have a different low-temperature asymptotic series.

Equation~\eqref{mshmu} for~$m_s$ can be expressed in terms of the independent variables $h\ge0$
and $0\le n_c\le1$ by taking into account how the regions~$\cB_i$, $\cT$ and $\cW_i$ transform
under the (non-invertible) mapping~$(h,\mu)\mapsto(h,n_c)$ determined by Eq.~\eqref{nc0}
(cf.~Fig.~\ref{fig.hmuhnc} right). To begin with, it is obvious that the wedge~$\cW_0$ collapses
into the line~$n_c=0$. Similarly, the wedge~$\cW_1$ is transformed into the horizontal half-line
$n_c=1$, $h>1/4$, while the vertical band $\cB_1$ goes into the segment $n_c=1$, $h<1/4$. On the
other hand, the triangle $\cT$ is mapped into the bounded region to the left of the
parabola~$h=h_0(n_c)$, where
\[
  h_0(n_c)=\vep(n_c/2)=\frac{n_c}4\,(2-n_c)\,.
\]
Indeed, in the triangle $\cT$ we have
\[
  n_c=2x_0(2\mu)\iff \vep(n_c/2)=2\mu > h\,.
\]
Likewise, the oblique band~$\cB_0$ is transformed into the unbounded region to the right of the
parabola~$h=h_0(n_c)$, since in this band
\[
  n_c=2x_0(\mu +\tfrac h{2})\iff \vep(n_c/2)=\mu +\frac h{2}< h\,.
\]
From these considerations it readily follows that the zero-temperature magnetization is expressed
in terms of the variables~$(h,n_c)$ by
\begin{equation}\label{ms0}
  m_s=
  \cases{
    x_0(h),& $0\le h\le h_0(n_c)$\\
    \frac12\,n_c,& $h\ge h_0(n_c)$
  }
\end{equation}
(cf.~Ref.~\cite{Ka92b}). Note that $m_s$ should be continuous everywhere in $(h,n_c)$ space, since
it is continuous when expressed in terms of the variables~$(h,\mu)$. In particular, the previous
expression for the critical magnetic field~$h_0$ is recovered by imposing the continuity of $m_s$
across the parabola $h=h_0(n_c)$.

From Eqs.~\eqref{mshmu}-\eqref{nc0} it is straightforward to compute the zero-temperature magnetic
and charge susceptibilities by differentiation. To begin with, the magnetic susceptibility
vanishes in the wedges~$\cW_0$ and~$\cW_1$. On the other hand, in the region~$\cB_1\cup \cT$
(including the segment~$\mu=1/8$, $h<1/4$) we have
\[
  \chi_s=\pdf{m_s}h=x_0'(h)=\frac1{1-2x_0(h)}= \frac{1}{1-2m_s}\,,\qquad (h,\mu)\in\cB_1\cup\cT\,.
\]
Likewise, in the band~$\cB_0$ the magnetic susceptibility is given by
\[
  \chi_s=\frac12\,x_0'\bigl(\mu +\tfrac h{2}\bigr)=
  \frac{1}{2(1-2m_s)}=\frac{1}{2(1-n_c)}\,,\qquad (h,\mu)\in\cB_0\,.
\]
A similar analysis for the charge susceptibility~$\chi_c\equiv\pd n_c/\pd\mu_c=\pd n_c/\pd\mu$
yields the result
\[
  \chi_c=
  \cases{
    0,& $(h,\mu)\in \cB_1\cup \cW_0\cup \cW_1$\\
    \dfrac{4}{1-n_c}\,,&$(h,\mu)\in \cT$\\
    \dfrac{2}{1-n_c}\,,& $(h,\mu)\in B_{0}$\,.
  }
\]
(In fact, $\chi_c$ vanishes also on the segment $h=1/4$, $\mu>1/8$.) Comparing with the previous
expressions for $m_s$ and~$n_c$ at zero temperature, we deduce that~$\chi_s$ diverges on the
half-lines~$\{h=1/4,\mu\ge 1/8\}$ and~$\{\mu+h/2=1/4,\mu\le 1/8\}$, while~$\chi_c$ is divergent on
the half-lines~$\{\mu=1/8,h<1/4\}$ and~$\{\mu+h/2=1/4,\mu\le 1/8\}$ . This is a well-known fact
(see, e.g., Ref.~\cite{KK95}). More surprising is the behavior of the magnetic and charge
susceptibilities at the boundaries of the~$\su(1|1)$ phase with the $\su(1|2)$ phase and the
vacuum. Indeed, these functions present jump discontinuities on the segment
$\{\mu=h/2,0\le h\le 1/4\}$ and the half-line~$\{\mu+h/2=0,h\ge0\}$. The discontinuity on the
latter segment (which, to the best of our knowledge, had not been previously pointed out in the
literature) is particularly interesting, since it is due to the fact that~$(1-2m_s)\chi_s$
and~$(1-n_c)\chi_c$ are different constants on each side of this segment. It is also worth
mentioning that $\chi_c=4\chi_s$ on the union of the half-planes $\mu<h/2$ and~$h>1/4$. Note,
finally, that our formulas for~$\chi_s$ and~$\chi_m$ agree with those of Ref.~\cite{KK95} in the
triangle~$\cT$, which is the only region considered in the latter reference\footnote{Note that the
  magnetic field and the magnetic moment in the Ref.~\cite{KK95} are respectively $h/2$ and~$2m_s$
  in our notation. Indeed, in the latter reference the magnetic field interaction in the
  Hamiltonian is taken as $h(n^1-n^2)$, while the magnetization is defined as
  $\langle n^1-n^2\rangle$.}. In any case, the dependence of~$\chi_s$ (resp.~$\chi_c$) exclusively
on~$m_s$ (resp.~on $n_c$) at zero temperature is a manifestation of the spin-charge separation
characteristic of the $t$-$J$ model.

It is also straightforward to express the zero-temperature susceptibilities as functions of the
variables~$(h,n_c)$. The key fact in this respect is that the segment~$\mu=h/2$, $0\le h\le 1/4$
is mapped to the arc of the parabola $n_c=2x_0(h)\equiv1-\sqrt{1-4h}$ (or, equivalently,
$h=h_0(n_c)$) with~$0\le h\le 1/4$. The susceptibilities are then given by
\[
  \chi_s=
  \cases{
    0,& $(h,n_c)\in \cW_0'\cup \cW_1'$\\
    \dfrac{1}{1-2m_s}\equiv
    \dfrac1{\sqrt{1-4h}},& $(h,n_c)\in \cB_{1}'\cup \cT'$\\
    \dfrac{1}{2(1-2m_s)}\equiv\dfrac1{2(1-n_c)},& $(h,n_c)\in \cB_{0}'$
  }
\]
and
\[
  \chi_c=
  \cases{
    0\,,& $(h,n_c)\in \cB_1'\cup \cW_0'\cup \cW_1'$\\
    \dfrac4{1-n_c}\,,& $(h,n_c)\in \cT'$\\
    \dfrac2{1-n_c}\,,& $(h,n_c)\in\ \cB_{0}'$\,.
 }
\]
In particular, we see that~$\chi_c=4\chi_s$ in the infinite region~$\sqrt{1-4h}<1-n_c$.

\section{Asymptotic series for the free energy and criticality}\label{sec.FE}

Starting from the exact formula~\eqref{fKY}, in this section we shall derive the complete
asymptotic series of the free energy of the~$\su(2)$ KY model\footnote{From now on, by ``$\su(m)$
  KY model'' we shall understand the full Hamiltonian~\eqref{HKYmu}, whose free energy (in the
  thermodynamic limit) is therefore the grand potential of the original KY Hamiltonian~$H_0$ in
  Eq.~\eqref{tJSUSY}.} at~$T=0$ for arbitrary values of the parameters $h$ and $\mu$. We shall
also use this asymptotic series to analyze the model's criticality properties and the
low-temperature behavior of its main thermodynamic functions.

\subsection{Wedges $\cW_0$ and $\cW_1$}
To begin with, it is straightforward to show that in the wedges~$\cW_{0,1}$ the free
energy is exponentially small in~$\be$ as $T\to 0$. Indeed, we can rewrite Eq.~\eqref{fKY} as
\begin{equation}\label{f2}
  f(T)=-\mu_c-\frac h2
  -2T\int_0^{\mathrlap{1/2}}\en\log\Bigl[\tb+\sqrt{{\tb}^2+\e^{-\be(h-\vep)}-\e^{-\be
      h}}\,\,\Bigr]\,\diff x\,,
\end{equation}
with
\begin{equation}\label{tb}
  \tb(x)\equiv \e^{-\frac{\be h}2}b(x)=\frac12\big[1+\e^{-\be(\mu+h/2-\vep(x))}+\e^{-\be h}\big].
\end{equation}
Clearly all the exponents in the previous formula for~$f$ are strictly negative in the
region~$\cW_1$, so that~$f(0)=-\mu_c-h/2$. Taking in to account that~$\vep(x)\le\vep(1/2)=1/4$ we
easily obtain
\[
|f(T)-f(0)|\le T\log\bigl[a+\sqrt{a^2+\e^{-\be(h-1/4)}}\,\bigr],
\]
with $a\equiv\tb(1/2)>1/2$. From the elementary inequality~$\sqrt{a^2+x}\le a+x/(2a)$
(where $x>0$) it then follows that
\[
  a+\sqrt{a^2+\e^{-\be(h-1/4)}}\le 2a+\e^{-\be(h-1/4)}\,,
\]
which easily yields the estimate
\begin{equation*}%\label{W1bound}
  |f(T)-f(0)|=\Or\Bigl(T\e^{-\be\min\bigl(h-\frac 14,\mu+\frac h2-\frac 14\bigr)}\Bigr),\en
  (h,\mu)\in \cW_1\,.
\end{equation*}
A similar analysis in the region~$\cW_0$ shows that
\begin{equation*}%\label{W0bound}
  |f(T)-f(0)|=\Or\Bigl(T\e^{-\be|\mu+\frac h2|}\Bigr),\quad
  (h,\mu)\in \cW_0\,.
\end{equation*}
Recall that at low temperatures the free energy per unit length of a $(1+1)$-dimensional CFT (in
natural units $\hbar=k_{\mathrm B}=1$) behaves as~\cite{BCN86,Af86}
\begin{equation}\label{fCFT}
f(T)\simeq f(0)-\frac{\pi c T^2}{6 v}\,,
\end{equation}
where $c$ is the central charge and $v$ is the Fermi velocity (effective speed of light). From the
previous estimates for~$f(T)-f(0)$ at low temperatures it then follows that the $\su(2)$ KY model
is \emph{not} critical when~$(h,\mu)$ lies on the wedges~$\cW_0$ and~$\cW_1$. In fact, the
exponentially small bounds in~$\be$ for $|f(T)-f(0)|$ found above show that the spin $1/2$ KY
model is gapped on the wedges~$\cW_{0,1}$, with energy gap given by~$|\mu+h/2|$ in~$\cW_0$
and~$\min(h-1/4,\mu+h/2-1/4)$ in~$\cW_1$.

\subsection{Vertical band $\cB_1$}
Splitting the integration interval in Eq.~\eqref{f2} into $[0,x_0(h)]$
and $[x_0(h),1/2]$, and setting
\begin{equation}
  \hb(x)\equiv \e^{-\frac\be2\vep(x)}b(x)
  =\frac12\bigg[\e^{-\frac\be2(\vep-h)}+\e^{-\frac\be2(\vep+h)}+\e^{-\be(\mu-\frac{\vep}2)}\bigg],
\end{equation}
we can write
\begin{eqnarray}
  f(T)=f_0
  &-2T\int_0^{x_0(h)}\kern-1em\log\Bigl[\tb+\sqrt{{\tb}^2+\e^{-\be(h-\vep)}-\e^{-\be
      h}}\,\,\Bigr]\,\diff x\nonumber\\
  &-2T\int^{1/2}_{\mathrlap{x_0(h)}}\log\Bigl[\hb+\sqrt{\hb^2+1-\e^{-\be
  \vep}}\,\,\Bigr]\,\diff x\,,
  \label{f3}
\end{eqnarray}
where
\begin{equation}
  f_0=-\mu_c-hx_0(h)-\int_{\mathrlap{x_0(h)}}^{1/2}\vep(x)\diff x.\label{f0B1}
\end{equation}
When~$(h,\mu)\in \cB_1$ all the exponents in the previous formulas for~$\tb$ and $\hb$ are negative
for~$0\le x<x_0(h)$ and~$x_0(h)<x\le 1/2$, respectively. Thus both integrals in Eq.~\eqref{f3}
vanish at~$T=0$, and hence~$f_0=f(0)$. Furthermore, we have
\begin{eqnarray}
  &\int_0^{\mathrlap{x_0(h)}}\quad\log\Bigl[\tb+\sqrt{{\tb}^2+\e^{-\be(h-\vep)}-\e^{-\be
    h}}\,\,\Bigr]\,\diff
    x=I_1
  +\Or(\e^{-\be\min(h,\mu-\frac h2)}),\label{I1def}\\
  &\int^{1/2}_{\mathrlap{x_0(h)}}\log\Bigl[\hb+\sqrt{\hb^2+1-\e^{-\be
    \vep}}\,\,\Bigr]\,\diff x
  =I_2+\Or(\e^{-\be\min(h,\mu-\frac 18)}),
    \label{I2def}
\end{eqnarray}
where
\begin{eqnarray}
  I_1&\equiv\int_0^{\mathrlap{x_0(h)}}\quad\log\Bigl[\tfrac12+\sqrt{\tfrac14+\e^{-\be(h-\vep)}}\,
       \Bigr]\diff x\,,\label{I1def2}\\
  I_2&\equiv\int^{1/2}_{\mathrlap{x_0(h)}}\log\Bigl[\tfrac12\,\e^{-\frac\be2(\vep-h)}+\sqrt{1+\tfrac14\,
       \e^{-\be(\vep-h)}}\,\,\Bigr]\,\diff x
       \label{I2def2}
\end{eqnarray}
(cf.~\ref{app.A}). We shall next derive the full asymptotic series of the integrals~$I_{k}$ in
powers of~$T$. To this end, let us perform in the integral~$I_1$ the change of
variable~$y=\be(h-\vep(x))$, or equivalently
\begin{equation}\label{xychangem}
  x=x_0(h-Ty)\,,
\end{equation}
obtaining
\[
  I_1=T\int_0^{\mathrlap{\be h}}\log\Bigl[\tfrac12+\sqrt{\tfrac14+\e^{-y}}\,\Bigr]x_0'(h-Ty)\,\diff
  y\,.
\]
We next expand the last term in the previous equation around~$T=0$ taking into account the
identity $x_0'=(1-2x_0)^{-1}$, with the result
\[
  x_0'(h-Ty)=\sum_{l=0}^{\infty}a_l(h)(-Ty)^{l}\,,
\]
where
\begin{equation}\label{cl}
  a_l(s)=\frac{2^l(2l-1)!!}{l!\,{[1-2x_0(s)]}^{2l+1}}
  =\frac{2^l(2l-1)!!}{l!\,{(1-4s)}^{l+\frac12}}
\end{equation}
and $(-1)!!\equiv1$. As shown in~\ref{app.B}, the asymptotic series of~$I_1$ is then given by
\begin{equation}\label{I1asymp}
  I_1\sim\sum_{l=0}^\infty (-1)^l\,a_l(h)\,T^{l+1}\!\!\int_0^\infty
  y^l\log\Bigl[\tfrac12+\sqrt{\tfrac14+\e^{-y}}\,\Bigr]\diff y\,.
\end{equation}

Consider next the integral~$I_2$. Since the natural change of variable
\begin{equation}\label{xychange}
  x=x_0(h+Ty)
\end{equation}
is singular at the endpoint~$y=\be(1/4-h)$, we first subdivide the integration range into the
intervals~$x_0(h)\le x\le x_0(\frac 18+\frac h{2})$ and
$x_0(\frac 18+\frac h{2})\le x\le 1/2$. The integral over the second interval is clearly
$\Or(\e^{-\frac\be4(\frac 14-h)})$, so that
\begin{equation}
  I_2=\int_{x_0(h)}^{x_0(\frac 18+\frac h{2})}\kern-.5em
  \log\Bigl[\tfrac12\,\e^{-\frac\be2(\vep-h)}+\sqrt{1+\tfrac14\,\e^{-\be(\vep-h)}}\,\,\Bigr]\diff
  x+\Or(\e^{-\frac\be4(\frac 14-h)})\,.\label{I2new}
\end{equation}     
Performing the change of variable~\eqref{xychange} in the integral in Eq.~\eqref{I2new} and
proceeding as before we obtain
\begin{equation}\label{asympI2}
  I_2\sim \sum_{l=0}^\infty a_l(h)\, T^{l+1}\!\!\int_0^{\mathrlap{\infty}}
  y^l\log\Bigl[\tfrac12\,\e^{-y/2}+\sqrt{1+\tfrac14\,\e^{-y}}\,\Bigr]\diff y.
\end{equation}
From Eqs.~\eqref{f3}--\eqref{I2def} and~\eqref{I1asymp}-\eqref{asympI2} we finally obtain the
asymptotic series of the free energy per site in the open band~$\cB_1$:
\begin{equation}
  \label{fasyB1}
  f(T)-f(0)\sim -2\sum_{l=0}^\infty a_l(h)\,\mathcal I_l\,T^{l+2},\quad
  (h,\mu)\in \cB_1\,,
\end{equation}
where
\begin{equation}\label{cIl}
  \fl
  \cI_l=\int_0^{\mathrlap{\infty}}y^l\Big\{\log\Bigl[\tfrac12\,\e^{-y/2}+\sqrt{1+\tfrac14\,\e^{-y}}\,
  \Bigr]+(-1)^l\log\Bigl[\tfrac12+\sqrt{\tfrac14+\e^{-y}}\,\Bigr]\Big\}\diff y.
\end{equation}
As explained in~\ref{app.ints}, the latter integrals can be expressed in several alternative ways,
to wit
\begin{equation}
  \fl
  \cI_l=\frac1{2(l+1)}\int_{-\infty}^\infty\kern -9pt
  x^{l+1}\left[\frac1{\sqrt{1+4\e^x}}-\th(-x)\right]\diff
  x
  =\frac1{(l+1)(l+2)}\int_{-\infty}^\infty \frac{x^{l+2}\,\e^x}{(1+4\e^x)^{3/2}}\,\diff x,
  \label{Ilalt}
\end{equation}
where $\th(t)=(1+\mathop{\rm sgn} t)/2$ is Heaviside's step function. The integrals~$\cI_l$ can
actually be computed in closed form for low values of~$l$, namely
\[
  \cI_0=\frac{\pi^2}6\,,\quad \cI_1=\ze(3)\,,\quad \cI_2=\frac{\pi^4}{10}\,,\quad
  \cI_3=2\big[\pi^2\ze(3)+9\ze(5)\big],
\]
where~$\ze(z)$ denotes Riemann's zeta function. We thus obtain the low temperature expansion
\begin{equation}\label{fT4B1}
  \fl
  f(T)-f(0)=-\frac{\pi^2T^2}{3(1-4h)^{1/2}}-\frac{4\ze(3)T^3}{(1-4h)^{3/2}}
  -\frac{6\pi^4T^4}{5(1-4h)^{5/2}}+\Or(T^5),\quad
  (h,\mu)\in \cB_1.
\end{equation}

Equation~\eqref{fT4B1} strongly suggests that the~$\su(2)$ KY model is critical in the vertical
band~$\cB_1$. To ascertain this fact and compute the central charge, however, we first need to
determine the Fermi velocity~$v$ of the low-energy excitations above the ground state. To this
end, recall first of all that in the limit~$N\to\infty$ the ground state energy of the spin~$1/2$
KY model is approximately given by
\begin{equation}\label{uGS}
  E_0 \simeq \sum_{k=1}^{Nm_s}\big(\vep(x_k)-h\big)
  +\sum_{k=1}^{\mathclap{Nn_c/2}}\big(\vep(x_k)-2\mu\big)\,,\qquad x_k\equiv k/N\,,
\end{equation}
while its momentum (mod.~$2\pi$) can be written as
\begin{equation}\label{PGS}
  P \simeq 2\pi\sum_{k=1}^{\mathclap{Nm_s}}x_k
  +2\pi\sum_{k=1}^{\mathclap{Nn_c/2}}x_k
\end{equation}
(see~\cite{BBCFG19} and, e.g.,~\cite{HHTBP92,BBS08}). As shown in Section~\ref{sec.T0}, in the
vertical band~$\cB_1$ we have
\[
  m_s=x_0(h)\,,\qquad n_c=1\,,
\]
and hence
\[
  \fl
  E_0 \simeq \sum_{k=1}^{\mathclap{Nx_0(h)}}\big(\vep(x_k)-h\big)
  +\sum_{k=1}^{N/2}\big(\vep(x_k)-2\mu\big)\,,
  \qquad P\simeq2\pi \sum_{k=1}^{\mathclap{Nx_0(h)}}x_k+2\pi\sum_{k=1}^{\mathclap{N/2}}x_k\,.
\]
Low-energy excitations above the ground state are obtained by adding a ``mode'' with $k=Nx_0(h)+1$
or removing one with~$k=Nx_0(h)-1$. The energy of these excitations is thus~$E_0+\De E$, with
\[
  \fl \De E=\pm\big[\vep\bigl(x_0(h\bigr)\pm\tfrac1N)-h\big]
  \simeq\pm\bigg[\vep\bigl(x_0(h)\bigr)\pm\frac{\vep'\bigl(x_0(h)\bigr)}N-h\bigg] =
  \frac{\vep'\bigl(x_0(h)\bigr)}N\,,
\]
On the other hand, the momentum carried by the mode added (respectively removed) is
\[
  p=2\pi x_k\equiv p_0\pm\De p\,,
\]
where $p_0=2\pi x_0(h)$ is the Fermi momentum and~$\De p=2\pi/N$. The
Fermi velocity of the low-energy excitations is therefore given by
\begin{equation}\label{vB1}
  v=\frac{\De E}{\De p}=\frac{\vep'\bigl(x_0(h)\bigr)}{2\pi}
  =\frac{1-2x_0(h)}{2\pi}=\frac{\sqrt{1-4h}}{2\pi}\,.
\end{equation}
From Eqs.~\eqref{fT4B1} and~\eqref{vB1} it then follows that in this case the model is
critical\footnote{It is also important to mention in this respect that the ground state of the
  spin~$1/2$ KY model has finite degeneracy (at most $4$)~\cite{BBS08}.} with central charge
$c=1$. It should also be noted that in the limit~$T\to0$ the only contribution to the integrals
$I_{1,2}$ in Eqs.~\eqref{I1def}-\eqref{I2def}, in terms of which
\[
  f-f(0)\sim-2T(I_1+I_2)\,,
\]
comes from an arbitrarily small neighborhood of the point~$x_0(h)=p_0/(2\pi)$ up to exponentially
small terms in~$\be$. We shall express this relationship by saying that these integrals are
critical at $x=x_0(h)$. Thus the Fermi velocity~\eqref{vB1} is proportional to the derivative of
the dispersion relation~$\vep(x)$ at the unique critical point of the integrals~$I_{1,2}$.

\subsection{Oblique band~$\cB_0$}
In this case Eq.~\eqref{f3} becomes
\begin{eqnarray}
  f(T)=f_0
  &-2T\int_0^{x_0(\frac h{2}+\mu)}\kern-1em\log\Bigl[\tb+\sqrt{{\tb}^2+\e^{-\be(h-\vep)}-\e^{-\be
    h}}\,\,\Bigr]\diff x\nonumber\\
  &-2T\int^{1/2}_{x_0(\frac h{2}+\mu)}\kern-1em\log\Bigl[\bb+\sqrt{\bb^2+\e^{-\be
    (\vep-2\mu)}-\e^{-2\be(\vep-\mu)}}\,\,\Bigr]\diff x\,,
  \label{fB0}
\end{eqnarray}
where now
\begin{equation}\label{f0B0}
  f_0=-\mu_c-h\,x_0\bigl(\tfrac h2+\mu\bigr)-2\int_{\mathrlap{x_0(\frac h{2}+\mu)}}^{1/2}\,\big(\vep(x)-\mu\big)\,\diff x
\end{equation}
and
\begin{equation}
  \bb(x)\equiv \e^{\be(\mu-\vep(x))}\,b(x)
  =\frac12\,\Big[1+\e^{-\be(\vep-\frac h2-\mu)}+\e^{-\be(\vep+\frac h2-\mu)}\Big].
  \label{bbar}
\end{equation}
Again, all the exponents appearing in Eq.~\eqref{fB0} are nonpositive, so that~$f_0=f(0)$.
Moreover, proceeding as above we obtain the estimates
\begin{eqnarray}
  \fl
  &\int_0^{x_0(\frac h{2}+\mu)}\kern-1em\log\Bigl[\tb+\sqrt{{\tb}^2+\e^{-\be(h-\vep)}-\e^{-\be
    h}}\,\Bigr]\diff x=I_3
    +\Or(\e^{-\be\min(h,\frac h2-\mu)}),\label{I3est}\\
  \fl
  &\int^{1/2}_{x_0(\frac h{2}+\mu)}\kern-1em\log\Bigl[\bb+\sqrt{\bb^2+\e^{-\be
    (\vep-2\mu)}-\e^{-2\be(\vep-\mu)}}\,\Bigr]\diff x=I_4+\Or(\e^{-\be\min(h,\frac h2-\mu)}),
    \label{I4est}
\end{eqnarray}
where
\begin{eqnarray}
  \fl
  I_3&\equiv\int_0^{x_0(\frac h2+\mu)}\kern-1em\log\Bigl[1+\e^{-\be(\frac h2+\mu-\vep)}\,\Bigr]\diff
       x\,,\label{I3def}\\
  \fl
  I_4&\equiv\int^{1/2}_{x_0(\frac h2+\mu)}\kern-.9em\log\Bigl[1+\e^{-\be(\vep-\frac
       h2-\mu)}\,\Bigr]\,
       \diff x=\int^{x_0(\frac18+\frac h{4}+\frac\mu{2})}_{x_0(\frac h2+\mu)}\kern-.3em
       \log\Bigl[1+\e^{-\be(\vep-\frac h2-\mu)}\,\Bigr]\,\diff x\nonumber\\
  \fl
     &\hphantom{\int^{1/2}_{x_0(\frac h2+\mu)}\kern-.9em\log\Bigl[1+\e^{-\be(\vep-\frac
       h2-\mu)}\,\Bigr]\,\diff x=\int^{x_0(\frac18+\frac h{4}+\frac\mu{2})}_{x_0(\frac
       h2+\mu)}\kern-1em\log\Bigl[1+\e^{-\be\vep-h/2-\mu}}+\Or(\e^{-\frac\be2(\frac 14-\frac h2-\mu)})\,.
       \label{I4def}
\end{eqnarray}
The asymptotic series of the integral~$I_3$ is obtained as above through the change of
variable~$y=\be\bigl(\tfrac h2+\mu-\vep(x)\bigr)$, namely
\begin{eqnarray}
  I_3&\sim\sum_{l=0}^\infty (-1)^la_l\bigl(\tfrac h{2}+\mu\bigr)\,
       T^{l+1}\!\!\int_0^{\mathrlap{\infty}}y^l\log(1+\e^{-y})\diff y\nonumber\\
     &=\sum_{l=0}^\infty (-1)^la_l\bigl(\tfrac h{2}+\mu\bigr)\,l!\big(1-2^{-l-1}\big)\ze(l+2)\,T^{l+1}.
    \label{I3series}
\end{eqnarray}
Likewise, performing the analogous change of variable~$y=\be\bigl(\vep(x)-\tfrac h2-\mu\bigr)$ in
the RHS of Eq.~\eqref{I4def} we obtain the asymptotic series
\begin{equation}\label{I4series}
  I_4\sim\sum_{l=0}^\infty a_l\bigl(\tfrac h{2}+\mu)\,l!\big(1-2^{-l-1}\big)\ze(l+2)\,T^{l+1}.
\end{equation}
Combining Eqs.~\eqref{I3series}-\eqref{I4series} we finally arrive at the following asymptotic
series for the free energy per site in the oblique band~$\cB_0$:
\begin{equation}\label{fasyB0}
  f(T)-f(0)
  \sim-2\sum_{l=0}^\infty\frac{(2^{2l+1}-1)(4l-1)!!}{\big[1-2(h+2\mu)\big]^{2l+\frac12}}\,\ze(2l+2)\,
  T^{2l+2}.
\end{equation}
In particular, the first few terms in the latter series are explicitly given by\footnote{The
  coefficients~$\ze(2l+2)$ can be expressed as
  \[
    \ze(2l+2)=\frac{(2\pi)^{2l+2}}{2(2l+2)!}\,|B_{2l+2}|\,,
  \]
  where~$B_k$ is the $k$-th Bernoulli number defined by
    $x/(\e^x-1)=\smash{\sum\limits_{k=0}^\infty} B_k\,\dfrac{x^k}{k!}$\,.}
\[
  \fl
  f(T)-f(0)=-\frac{\pi^2T^2}{3\big[1-2(h+2\mu)\big]^{1/2}}\\
  -\frac{7\pi^4T^4}{15\big[1-2(h+2\mu)\big]^{5/2}}+\Or(T^6),\quad
  (h,\mu)\in \cB_0.
\]

As before, the previous asymptotic expansion indicates that the model is critical when~$(h,\mu)$
lie in the oblique band~$\cB_0$. In this case we have
\[
  m_s=\frac{n_c}2=x_0\bigl(\mu+\tfrac h2\bigr)\,,
\]
so that  the ground state energy, momentum and Fermi momentum are given by
\[
  \fl
  E_0 \simeq 2\sum_{k=1}^{\mathclap{Nx_0(\mu+\frac h2)}}\big(\vep(x_k)-\mu-\tfrac h2\big)\,,\qquad
  P\simeq2\cdot2\pi\sum_{k=1}^{\mathclap{Nx_0(\mu+\frac h2)}}x_k\,,
  \qquad p_0=4\pi x_0\bigl(\mu+\tfrac h2\bigr)\,,
\]
and therefore the low-energy excitations satisfy
\[
  \De E=\frac2N\,\vep'\bigl(x_0(\mu+\tfrac h2)\bigr)\,,\qquad \De p=\frac{4\pi}N\,.
\]
We conclude that the Fermi velocity is given in this case by
\[
  v=\frac{\vep'\bigl(x_0(\mu+\tfrac h2)\bigr)}{2\pi} =\frac{\sqrt{1-2(h+2\mu)}}{2\pi}\,,
\]
and thus the central charge is again~$c=1$. Note that, as in the previous case,
$x_0(\mu+\tfrac h2)$ is the unique critical point of the integrals~$I_{3,4}$ determining the
asymptotic expansion of~$f(T)-f(0)$ in the oblique band~$\cB_0$.

\subsection{Triangle~$\cT$}

We now have
\begin{eqnarray}
  \fl
  f(T)-f(0)
  =&-2T\int_0^{x_0(h)}\kern-.8em\log\Bigl[\tb+\sqrt{{\tb}^2+\e^{-\be(h-\vep)}-\e^{-\be
    h}}\,\,\Bigr]\diff x\nonumber\\
 &-2T\int^{x_0(2\mu)}_{x_0(h)}\kern-.8em\log\Bigl[\hb+\sqrt{\hb^2+1-\e^{-\be
    \vep}}\,\,\Bigr]\diff x\nonumber\\
  \fl
  &-2T\int_{x_0(2\mu)}^{1/2}\kern-.8em\log\Bigl[\bb+\sqrt{\bb^2+\e^{-\be
    (\vep-2\mu)}-\e^{-2\be(\vep-\mu)}}\,\,\Bigr]\diff x
    \label{fTtriangle}
\end{eqnarray}
with
\begin{equation}
  f(0)=-\mu_c-hx_0(h)-\int_{\mathrlap{x_0(h)}}^{x_0(2\mu)}\kern-.5em\vep(x)\,\diff x
  -2\int_{\mathrlap{x_0(2\mu)}}^{1/2}\,\big(\vep(x)-\mu\big)\,\diff x.\label{f0T}
\end{equation}
The first integral in Eq.~\eqref{fTtriangle} coincides with the LHS of Eq.~\eqref{I1def}, so that
its asymptotic series is given by Eq.~\eqref{I1asymp}. The last integral in Eq.~\eqref{fTtriangle}
differs from
\[
  \int_{\mathrlap{x_0(2\mu)}}^{x_0(\frac
    18+\mu)}\kern-.5em\log\Bigl[\tfrac12+\sqrt{\tfrac14+\e^{-\be(\vep-2\mu)}}\,\Bigr]\diff x
\]
by terms~$\Or(\e^{-\be(\frac 18-\mu)})$, and thus its asymptotic series can be computed performing
the change of variable~$y=\be(\vep(x)-2\mu)$ in the latter integral, with the result
\begin{equation}\label{asymp4}
  \sum_{l=0}^\infty a_l(2\mu)\,T^{l+1}\!\!\int_0^{\mathrlap{\infty}}
  y^l\log\Bigl[\tfrac12+\sqrt{\tfrac14+\e^{-y}}\,\Bigr]\diff y\,.
\end{equation}
Finally, the second integral in Eq.~\eqref{fTtriangle} is dominated by the
term~$\frac12\e^{-\be(\vep-h)}$ in~$\hb$ for~$x_0(h)\le x\le x_0(\frac h2+\mu)$, while in the
interval $[x_0(\frac h2+\mu),x_0(2\mu)]$ the dominant term is instead
$\frac12\e^{-\be(\mu-\frac{\vep}2)}$. More precisely, we have
\begin{eqnarray}
  \fl
  \int_{x_0(h)}^{x_0(\frac h2+\mu)}
  &\kern-.8em\log\Bigl[\hb+\sqrt{\hb^2+1-\e^{-\be
    \vep}}\,\,\Bigr]\diff x\nonumber\\
  \fl
  & =\int_{x_0(h)}^{x_0(\frac h{2}+\mu)}\kern-.8em\log\Bigl[
    \tfrac12\,\e^{-\frac\be2(\vep-h)}+\sqrt{1+\tfrac14\,\e^{-\be(\vep-h)}}\,\Bigr]\diff
    x+\Or(\e^{-\be\min(\frac12(\mu-\frac h2),h)})\label{asymp21}
\end{eqnarray}
and
\begin{eqnarray}
  \fl
 \int_{x_0(\frac h2+\mu)}^{x_0(2\mu)}&\kern-.8em\log\Bigl[\hb+\sqrt{\hb^2+1-\e^{-\be
                                        \vep}}\,\,\Bigr]\diff x\nonumber\\
  \fl
  &=\int_{x_0(\frac h{2}+\mu)}^{x_0(2\mu)}\kern-.8em\log\Bigl[
    \tfrac12\,\e^{-\frac\be2(2\mu-\vep)}+\sqrt{1+\tfrac14\,\e^{-\be(2\mu-\vep)}}\,\Bigr]\diff x
   +\Or(\e^{-\frac\be2(\mu-\frac h2)}).\label{asymp22}
\end{eqnarray}
Comparing with Eq.~\eqref{I2new} we conclude that the asymptotic series of the LHS of
Eq.~\eqref{asymp21} is given by Eq.~\eqref{asympI2}. On the other hand, the asymptotic series of
the LHS of Eq.~\eqref{asymp22} is easily derived performing the change of
variable~$y=\be(2\mu-\vep(x))$ in the RHS, with the result
\begin{equation}\label{asymp22new}
  \sum_{l=0}^\infty (-1)^la_l(2\mu)\,T^{l+1}\!\!\int_0^{\mathrlap{\infty}}
  y^l\log\Bigl[\tfrac12\,\e^{-y/2}+\sqrt{1+\tfrac14\,\e^{-y}}\,\Bigr]\diff y\,.
\end{equation}
Putting all of the above together we obtain the following asymptotic series for the free energy
per site in the triangle~$\cT$:
\begin{equation}\label{fasyT}
  f(T)-f(0)\sim-2\sum_{l=0}^\infty\Big[a_l(h)+(-1)^la_l(2\mu)\Big]
  \,\cI_l\,T^{l+2},
\end{equation}
where the integral $\cI_l$ is given by Eq.~\eqref{cIl}. In particular, in this case
\begin{equation}\label{psi}
  f(T)-f(0)\sim \psi(T,h)+\psi(-T,2\mu)\,,
\end{equation}
where
\begin{equation}
  \label{psiTh}
  \psi(T,h)=-2\sum_{l=0}^\infty a_l(h)\,\mathcal I_l\,T^{l+2}
\end{equation}
is the asymptotic series for $f(T)-f(0)$ in the region~$\cB_1$ (cf.~Eq.~\eqref{fasyB1}).

As in the previous two cases, the asymptotic behavior of the free energy near $T=0$ spelled out in
Eq.~\eqref{fasyT} is a strong indication that the model is critical when~$(h,\mu)$ belongs to the
triangle~$\cT$. To confirm this indication and compute the central charge, note first that in this
case
\[
  m_s=x_0(h)\,,\qquad n_c = 2x_0(2\mu)\,,
\]
and therefore the ground state energy and momentum are given by
\[
  \fl E_0 \simeq \sum_{k=1}^{\mathclap{Nx_0(h)}}\big(\vep(x_k)-h\big)
  +\sum_{k=1}^{\mathclap{Nx_0(2\mu)}}\big(\vep(x_k)-2\mu\big)\,, \qquad P\simeq
  2\pi\sum_{k=1}^{\mathclap{Nx_0(h)}}x_k +2\pi\sum_{k=1}^{\mathclap{Nx_0(2\mu)}}x_k\,.
\]
We thus have two types of low-energy excitations associated to spin and charge, with
$\De E=\vep'\bigl(x_0(h)\bigr)/N$ and $ \De E=\vep'\bigl(x_0(2\mu)\bigr)/N$ respectively for the
spin and charge excitations. Since in both cases~$\De p=2\pi/N$, the Fermi velocities of the spin
and charge excitations are respectively given by
\begin{equation}\label{vsvc}
  \fl
  v_s=\frac{\vep'\bigl(x_0(h)\bigr)}{2\pi}=\frac{\sqrt{1-4h}}{2\pi}\,,\qquad
  v_c=\frac{\vep'\bigl(x_0(2\mu)\bigr)}{2\pi}=\frac{\sqrt{1-8\mu}}{2\pi}\,.
\end{equation}
The leading term in the expansion~\eqref{fasyT} can therefore be expressed as
\begin{eqnarray}
  f(T)-f(0)&\simeq-2\big[a_0(h)+a_0(2\mu)]\cI_0
  T^2=-\frac{\pi^2T^2}3\,\bigg(\frac1{\sqrt{1-4h}}+\frac1{\sqrt{1-8\mu}}\,\bigg)\nonumber\\
  &=-\frac{\pi T^2}6\,\bigg(\frac1{v_s}+\frac1{v_c}
    \bigg)\,.
    \label{fTvsvc}
\end{eqnarray}
Hence both the charge and the spin sectors of the model are described at low energies by a CFT
with central charge $c=1$. This is indeed known to be the case, as first shown in
Ref.~\cite{Ka92b} using the asymptotic Bethe ansatz. Note finally that, as in the previous cases,
the points $x_0(h)$ and $x_0(\mu)$ appearing in Eq.~\eqref{vsvc} for the Fermi velocities are
nothing but the critical points of the integrals in the RHS of Eq.~\eqref{fTtriangle} for
$f(T)-f(0)$.

\subsection{Critical behavior on the boundaries}
The asymptotic behavior of the free energy on the boundaries of the five ground-state phases
$\cW_i$, $\cB_i$, $\cT$ can be analyzed in much the same way as above. In particular, it is not
difficult, and is certainly of interest, to examine the criticality properties of the model on
these boundaries. Consider, as an example, the segment $\mu=1/8$, $0<h<1/4$ separating the
triangle~$\cT$ ($\su(1|2)$ phase) from the vertical band~$\cB_1$ ($\su(2)$ phase). From
Eqs.~\eqref{fTtriangle} and~\eqref{asymp22} with~$\mu=1/8$ we readily obtain
\[
  f(T)-f(0)\sim-\frac{\pi T^2}{6v_s}-2T\int_{x_0(\frac h2+\frac18)}^{1/2}\kern-.5em
  \log\Bigl[
    \tfrac12\,\e^{-\frac\be2(\frac14-\vep)}+\sqrt{1+\tfrac14\,\e^{-\be(\frac14-\vep)}}\,\Bigr]\diff x\,,
\]
where the Fermi (spin) velocity~$v_s$ is given by Eq.~\eqref{vsvc}. Although the last integral is
critical at $x=1/2$, it is not asymptotic to~$-\pi T^2/(6v_c)$ as the Fermi velocity~$v_c$
vanishes at~$\mu=1/8$. In fact, performing the usual change of variable
\[
  y=\be(1/4-\vep(x))=\left(\frac12-x\right)^{\!2}
\]
we have
\begin{eqnarray*}
  \kern-1em
  2T\int_{x_0(\frac h2+\frac18)}^{1/2}\kern-.5em \log\Bigl[
    \tfrac12\,\e^{-\frac\be2(\frac14-\vep)}&+\sqrt{1+\tfrac14\,\e^{-\be(\frac14-\vep)}}\,\Bigr]\diff
    x\\
  &=T^{3/2}\int_0^{\frac\be2(\frac14-h)}\kern-1.4em \log\Bigl[
    \tfrac12\,\e^{-\frac y2}+\sqrt{1+\tfrac14\,\e^{-y}}\,\Bigr]\frac{\diff y}{\sqrt y}
  \sim \ka_c T^{3/2},
\end{eqnarray*}
with
\[
  \ka_c\equiv\int_0^\infty\log\Bigl[ \tfrac12\,\e^{-\frac
    y2}+\sqrt{1+\tfrac14\,\e^{-y}}\,\Bigr]\frac{\diff y}{\sqrt y} \simeq 1.2255036\,.
\]
Thus the first two nonvanishing terms in the asymptotic expansion of~$f(T)-f(0)$ on the
segment~$\mu=1/8$, $0<h<1/4$ are\footnote{In fact, it is straightforward to obtain the full
  asymptotic series
  \[
    f(T)-f(0)\sim-\ka_cT^{3/2}-2\sum_{l=0}^{\infty}a_l(h)\cI_l\,T^{l+2}\,.
  \]}
\[
  f(T)-f(0)\sim -\ka_c T^{3/2}-\frac{\pi T^2}{6v_s}\,.
\]
We conclude that the model is not critical on this segment, due to the term in the expansion
proportional to~$T^{3/2}$. However, the second term (proportional to $T^2$) can be interpreted as
signaling that on the spin sector the model is critical with central charge $c=1$. This conclusion
is borne out by the behavior of the ground state energy and momentum, which in this case are given
by
\[
  \fl E_0\simeq\sum_{k=1}^{\mathclap
    {Nx_0(h)}}\big(\vep(x_k)-h\big)+\sum_{k=1}^{N/2}\big(\vep(x_k)-\tfrac14\big)\,, \qquad P\simeq
  2\pi\sum_{k=1}^{\mathclap{Nx_0(h)}}x_k+2\pi\sum_{k=1}^{N/2}x_k\,.
\]
Since now $\vep'(1/2)=0$, the low energy excitations obtained by removing the mode with $k=N/2-1$
(or adding the one with $k=N/2+1$) carry an energy
\[
  \De E=\frac14-\vep(\tfrac12-\tfrac1N)=-\frac{\vep''(\tfrac12)}{2N^2}=\frac1{N^2}\,,
\]
which is \emph{quadratic} in~$\De p=2\pi/N$. These are in fact the excitations responsible for the
$T^{3/2}$ asymptotic behavior of~$f(T)-f(0)$ at low temperatures. On the other hand, exciting the
mode with $k=Nx_0(h)+1$ (or suppressing the one with $k=Nx_0(h)-1$) increases the energy by
\[
  \De E=\frac{\vep'(x_0(h))}N=v_s\De p\,.
\]
These excitations are therefore described by a CFT with Fermi velocity~$v_s$.

Proceeding in an analogous way we can determine the critical behavior (including, where
appropriate, the value of the central charge) in the remaining parts of the boundary. In general,
the vanishing of the spin or charge Fermi velocity implies that the model is not critical in the
corresponding sector (and, hence, as a whole), although it can still be critical in the other
sector provided that its Fermi velocity is nonzero. Since $v_s$ (respectively $v_c$) vanishes only
for~$h=1/4$ (respectively $\mu=1/8$), we conclude that the $\su(2)$ KY model should be critical at
the vertical segment~$h=0$, $0\le\mu<1/8$, non-critical (but critical in the spin sector) at the
point~$(0,1/8)$ and non-critical (in both sectors) at the other vertex $(1/4,1/8)$. This is indeed
confirmed by a detailed calculation (see Table~\ref{table.crit} for a summary of the results).
This calculation also shows that in all the non-critical parts of the boundary the model is
gapless, with $f(T)-f(0)$ growing as~$T^{3/2}$ at low temperatures.

\begin{table}[t]
  \centering
  \begin{tabular}{|c|c|c|}\hline
    Region & Central charge \\ \hline\hline
    $h=0,\en 0<\mu<\frac18$ & $1+1$\vru\\ \hline
    $h=0,\en \mu>\frac18$ & $1$\vru\\ \hline
    $\mu=\frac18,\en 0\le h<\frac14$ & $1$ (only spin sector)\vru\\ \hline
    $\mu=\frac h2,\en 0<h<\frac14$ & $\frac32$\vru\\ \hline
    $h=\frac 14,\en \mu>\frac18$ & Non-critical\vru\\ \hline
    $\mu=\frac14-\frac h2,\en h>\frac 14$ & Non-critical\vru\\ \hline
    $(0,0)$ & $1$\vru\\ \hline
    $(\frac14,\frac18)$ & Non-critical\vru\\ \hline
  \end{tabular}
  \caption{Critical behavior of the spin $1/2$ KY model on the boundaries of the ground-state
    phases~$\cW_i$, $\cB_i$, $\cT$. In all the non-critical regions (including the
    segment~$\mu=1/8$, $0\le h<1/4$) the leading term in the low-temperature asymptotic expansion
    of $f(T)-f(0)$ is proportional to~$T^{3/2}$ and the model is gapless.}\label{table.crit}
\end{table}

An interesting situation presents itself on the oblique segment~$\mu=h/2$, $0<h<1/4$. Indeed, by
Eq.~\eqref{fTtriangle} ---or \eqref{fB0}--- with~$\mu=h/2$ we have
\begin{eqnarray*}
  \fl
  f(T)-f(0)=
  -2T&\int_0^{x_0(h)}\kern-1em\log\Bigl[\tb+\sqrt{{\tb}^2+\e^{-\be(h-\vep)}-\e^{-\be
       h}}\,\,\Bigr]\diff x\\
     &-2T\int^{1/2}_{x_0(h)}\kern-1em\log\Bigl[\bb+\sqrt{\bb^2+\e^{-\be
       (\vep-h)}-\e^{-\be(2\vep-h)}}\,\,\Bigr]\diff x\equiv -2T(J_1+J_2)\,,
\end{eqnarray*}
where
\[
  \tb=\frac12\Big(1+\e^{-\be(h-\vep(x))}+\e^{-\be h}\Big)\,,\qquad
  \bb=\frac12\Big(1+\e^{-\be(\vep(x)-h)}+\e^{-\be \vep(x)}\Big)\,.
\]
Since both integrals are critical at~$x_0(h)\equiv x_0(2\mu)$, the Fermi velocity is expected to
be
\[
  v=\frac{\vep'\bigl(x_0(h)\bigr)}{2\pi}=\frac{\sqrt{1-4h}}{2\pi}>0\,.
\]
This is indeed the case, since the continuity of the ground state energy, momentum, magnetization
and charge density implies that when $h=2\mu$ we have
\[
  \fl m_s=\frac{n_c}2=x_0(h)\,,\quad
  E_0\simeq2\sum_{k=1}^{\mathclap{Nx_0(h)}}\big(\vep(x_k)-h\big)\,, \quad
  P\simeq2\cdot2\pi\sum_{k=1}^{\mathclap{Nx_0(h)}}x_k\,,
\]
and therefore the low-energy excitations satisfy
\[
  \De E=\frac2N\,\vep'\bigl(x_0(h)\bigr)\,,\qquad\De p=\frac{4\pi}N\,.
\]
Performing the changes of variable~$y=\be(h-\vep(x))$ and~$y=\be(\vep(x)-h)$ respectively in the
integrals $J_1$ and $J_2$, and proceeding as above, we easily obtain
\[
  J_{1,2}\sim-\frac{T}{2\pi v}\int_0^{\mathrlap{\infty}}\log\Bigl[\tfrac12(1+\e^{-y})
  +\sqrt{\tfrac14(1+\e^{-y})^2+\e^{-y}}\,\,\Bigr]\diff y=-\frac{\pi T}{16v}\,,
\]
and therefore
\[
  f(T)-f(0)\sim -\frac{\pi T^2}{4v}\,.
\]
Thus in the segment $\mu=h/2$, $0<h<1/4$ the low-temperature behavior of the model is described by
a single CFT with $c=3/2$. It is also interesting to note that, although both Fermi velocities are
nonzero in this case, this result cannot be obtained setting $v_s=v_c$ in Eq.~\eqref{fTvsvc} for
the triangle~$\cT$. The reason is of course that the terms~$\e^{-\be(\mu+h/2-\vep(x))}$ and
$\e^{-\be(\vep(x)-\mu-h/2)}$ appearing respectively in $\tb$ and $\bb$ are exponentially small
throughout their whole integration ranges~$[0,x_0(h)]$ and $[x_0(2\mu),1/2]$ (and can therefore be
discarded) only if $\mu$ is strictly greater than $h/2$.

\subsection{Discussion}
It should be noted that the expansions~\eqref{fasyB1}, \eqref{fasyB0} and~\eqref{fasyT} are all
true asymptotic series, i.e., their radius of convergence vanishes. Indeed, for
Eqs.~\eqref{fasyB1} and~\eqref{fasyT} this stems from the following bound on the integral~$\cI_l$
in Eq.~\eqref{cIl}:
\[
  \cI_l\ge\int_0^{\mathrlap{\infty}}\big[y^l(c\,\e^{-y/2}-\e^{-y})\big]\diff y=(2^{l+1}c-1)l!,
\]
where~$c\equiv\log(\frac{1+\sqrt5}2)$. To derive the latter bound simply observe that the function
\[
  \phi(y)\equiv\e^{y/2}\log\Bigl[\tfrac12\,\e^{-y/2}+\sqrt{1+\tfrac14\,\e^{-y}}\,\Bigr]
\]
is monotonically increasing on~$[0,\infty]$, so that~$\phi(y)\ge\phi(0)=c$. As
to~Eq.~\eqref{fasyB0}, from the elementary identity
\[
  \log(1+\e^{-y})\ge\log 2\cdot\e^{-y},\qquad y\ge0,
\]
it follows that
\[
  \int_0^{\mathrlap{\infty}}y^k\log\bigl(1+\e^{-y}\bigr)\diff
  y\ge\log2\int_0^{\mathrlap{\infty}}y^k\e^{-y}\diff y=k!\log 2\,.
\]
Our claim then follows from the fact that the coefficient of~$T^{2l+2}$ in
Eq.~\eqref{fasyB0} is proportional to
\[
  a_{2l}(\tfrac h{2}+\mu)\int_0^{\mathrlap{\infty}}y^{2l}\log\bigl(1+\e^{-y}\bigr)\diff y
\]
(cf.~\eqref{I3series} and its analog
\[
I_4\sim\sum_{l=0}^\infty a_l(\tfrac h{2}+\mu)\,T^{l+1}\int_0^{\mathrlap{\infty}}y^l\log(1+\e^{-y})\diff y
\]
for~$I_4$).
\begin{figure}[h]
  \includegraphics[width=.48\textwidth]{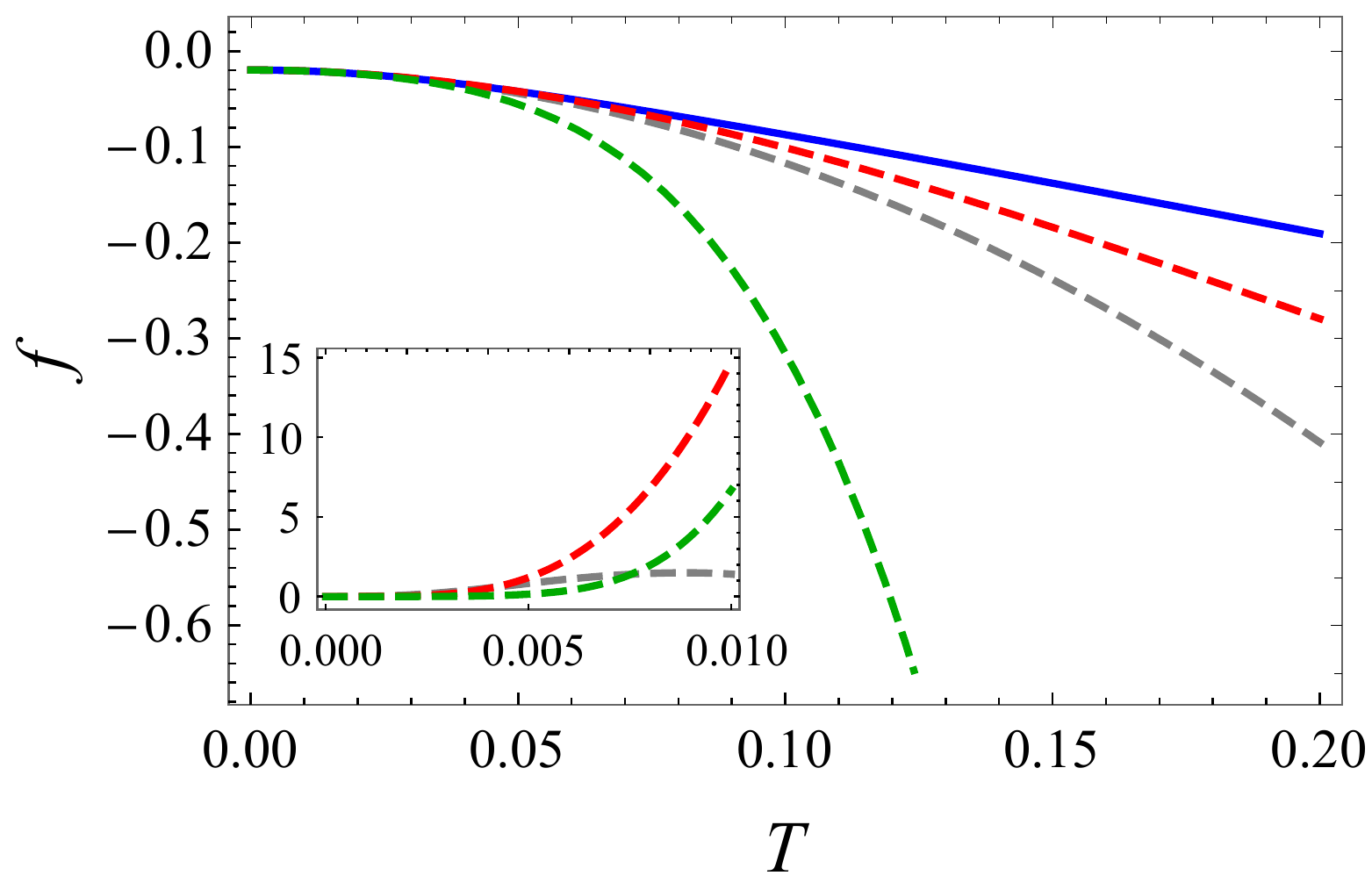}\hfill
  \includegraphics[width=.5\textwidth]{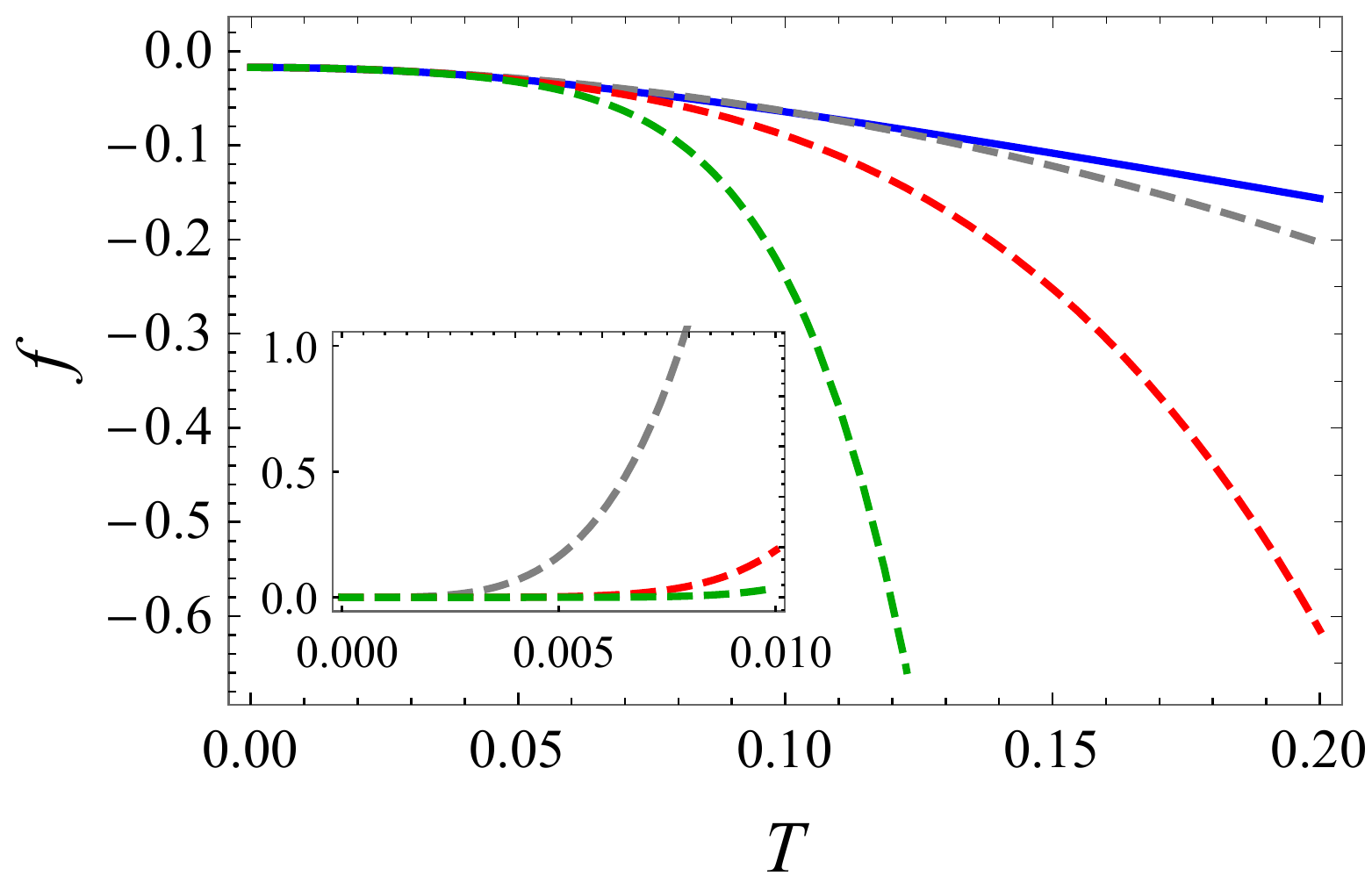}
  \caption{Left: comparison between $f$ (solid blue line) and its asymptotic
    expansions~\eqref{fasyT} with two, three and four terms (gray, red and green dashed lines,
    respectively) at the point~$h=\mu=1/12$ in the triangle~$\cT$. Right: analogous plot for the
    asymptotic expansion~\eqref{fasyB0} at the point $h=1/4$, $\mu=0$ in the oblique band~$\cB_0$.
    Insets: absolute value of the errors of the latter expansions (in units of~$10^{-6}$) in the
    smaller range~$0\le T\le 0.01$. In all plots, the unit of temperature and energy
    is~$K=2\pi^2t$.}
  \label{fig.fasy}
\end{figure}

In Fig.~\ref{fig.fasy} we compare the free energy per site numerically computed through
Eq.~\eqref{fKY} with its asymptotic expansions up to four terms at the points~$h=\mu=1/12$ in the
triangle~$\cT$ and $h=1/4,\mu=0$ in the oblique band~$\cB_0$ (cf.~Eqs.~\eqref{fasyT}
and~\eqref{fasyB0}). We see from this figure that the agreement between the exact value of~$f$ and
its expansions is quite good at sufficiently low temperature. To be more precise, from the
estimates for each of the integrals in Eq.~\eqref{fTtriangle} it can be checked that the
exponentially small terms discarded to obtain the asymptotic series~\eqref{fasyT} for $f$ in the
triangle~$\cT$ are $\Or(T\e^{-\be/48})$ at the point $h=\mu=1/12$. Thus the latter series should
not be expected to provide a good approximation for~$f$ unless $T\lesssim 0.02$. This is clearly
in agreement with the numerical results represented in Fig.~\ref{fig.fasy} (left). In particular,
from the inset in the latter figure it is apparent that the absolute value of the error of the
asymptotic expansions up to four terms does not exceed $1.5\times10^{-5}$ for $T\le 0.01$. A
similar remark can be made for the band~$\cB_0$ (see Fig.~\ref{fig.fasy}, right). Note also that,
although the absolute value of the error of the three asymptotic expansions considered diminishes
with their order at sufficiently low temperature (cf.~the insets in Fig.~\ref{fig.fasy}), this
need not be the case at higher temperatures. In fact, it is a well-known feature of divergent
asymptotic series that the optimum order varies with the range of the independent variable
considered.

\subsection{Comparison with the $\su(1|1)$ and~$\su(2)$ HS chains}

\begin{figure}[t]
  \includegraphics[width=.49\textwidth]{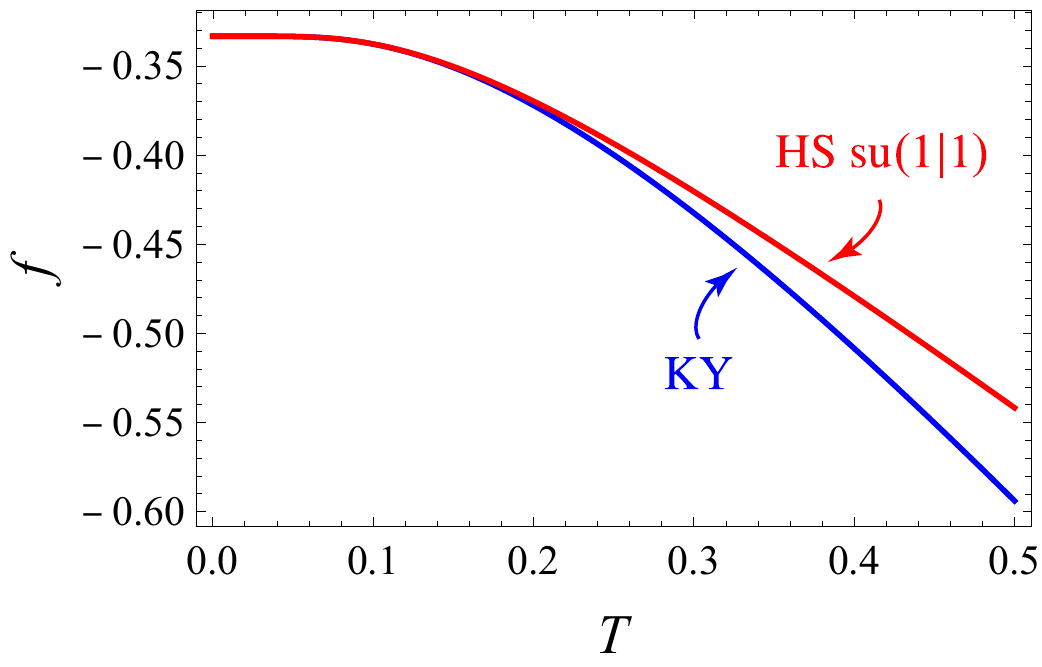}\hfill\includegraphics[width=.49\textwidth]{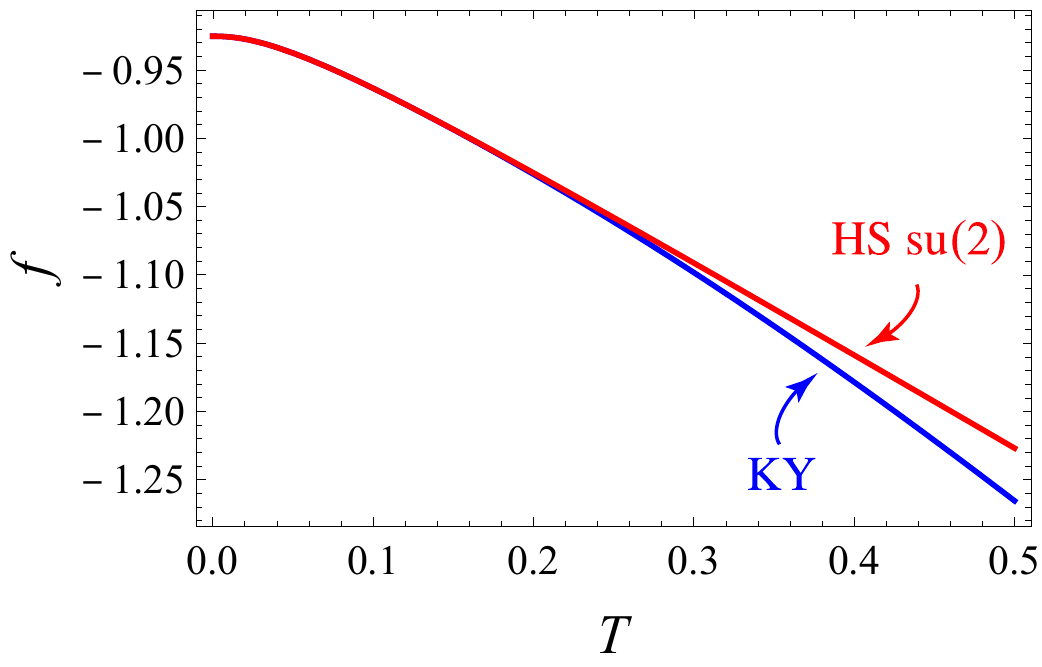}
  \caption{Left: comparison between the free energy of the supersymmetric~spin~$1/2$
    Kuramoto--Yokoyama model at the point $h=1$, $\mu=0$ in the band~$\cB_0$ and that of the
    (appropriately rescaled) $\su(1|1)$ HS chain~\eqref{H11} (cf.~Eq.~\eqref{f11resc}). Right:
    analogous comparison at the point $h=1/8$, $\mu=1$ in the band~$\cB_1$ with the
    (antiferromagnetic) rescaled $\su(2)$ HS chain. The unit of energy and temperature in both
    plots is $K=2\pi^2t$.}
  \label{fig.su112}
\end{figure}

It should be clear from the above results that the asymptotic series of the free energy per site
exhibits a different qualitative behavior in each of the regions $\cB_i$, $\cW_i$ and~$\cT$
identified in the previous section (cf.~Table~\ref{table.regs} and Fig.~\ref{fig.hmuhnc}, left).
Moreover, we shall next show that in the bands $\cB_{0}$ and~$\cB_1$, corresponding respectively
to the $\su(1|1)$ and~$\su(2)$ zero-temperature phases, the asymptotic series for~$f$ coincides
term by term with those for the free energy of the $\su(1|1)$ and~$\su(2)$ HS chains (with a
chemical potential and a magnetic field term, respectively). This does not mean that the
supersymmetric KY model is equivalent to the $\su(1|1)$ and~$\su(2)$ HS chains in these regions,
since their respective free energies differ by exponentially small terms in $\be$ which become
significant as $T$ increases (cf.~Fig.~\ref{fig.su112}).

Consider, to begin with, the $\su(1|1)$ HS chain with a chemical potential term, whose Hamiltonian
shall be taken in accordance with Ref.~\cite{CFGRT16} as
\begin{equation}
  \label{H11}
  \hat H^{(1|1)}=\frac{\pi^2}{N^2}\sum_{i<j}\frac{1-P_{ij}^{(1|1)}}{\sin^2(\pi(i-j)/N)}
  -\la\cF\,.
\end{equation}
In the oblique band~$\cB_0$, the number of ``down'' fermions at~$T=0$ is~$n^2=0$. We should then
compare~\eqref{H11} with the (spin chain version of) the Hamiltonian of the supersymmetric KY
model in the sector~$n^2=0$, given by
\begin{equation}\label{HB0}
  \hat H|_{\cN_2=0} = \frac1{2N^2}\sum_{i<j}\frac{1-P_{ij}^{(1|1)}}{\sin^2(\pi(i-j)/N)}
  -(\mu+\tfrac h2)\cN_1
\end{equation}
(cf.~Eq.~\eqref{hatH0H1}). We thus see that $\hat H^{(1|1)}=2\pi^2\hat H|_{\cN_2=0}$ provided
that~$\la=2\pi^2(\mu+h/2)$. The free energy of the Hamiltonian~$\hat H^{(1|1)}$ was computed
in~Ref.~\cite{CFGRT16}, namely
\[
  f^{(1|1)}(T)=-\frac{T}{\pi}\int_0^\pi\log\Bigl[1+\e^{-2\pi^2\be(\vep(\frac p{2\pi})-\frac h2-\mu)}\Bigr]\,\diff
  p\,.
\]
Taking into account the connection between $\hat H^{(1|1)}$ and $\hat H|_{\cN_2=0}$, we should
then expect that in the oblique band~$\cB_0$, and at sufficiently low temperature,
\begin{equation}\label{f11resc}
  \fl
  f(T)\simeq \frac1{2\pi^2}f^{(1|1)}(2\pi^2 T)=
  -2T\int_0^{1/2}\log\Bigl[1+\e^{-\be(\vep(x)-\frac h2-\mu)}\Bigr]\,\diff x\,.
\end{equation}
In fact, from Eqs.~\eqref{f0B0}-\eqref{I4est} it readily follows that
\begin{eqnarray*}
  f(T)&=f_0-2T(I_3+I_4)+\Or(\e^{-\be\min(h,\frac h2-\mu)})\\
      &=-2T\int_0^{1/2}\log\Bigl[1+\e^{-\be(\vep(x)-\frac h2-\mu)}\Bigr]\,\diff x
  +\Or(\e^{-\be\min(h,\frac h2-\mu)})\,.
\end{eqnarray*}
Thus the low-temperature asymptotic series of $f^{(1|1)}(2\pi^2T)/(2\pi^2)$ coincides term by term
with that of $f$ in the band~$\cB_0$, as claimed.

Similarly, in the vertical band~$\cB_1$ the number of bosons at~$T=0$ is~$n^0=0$. The (spin chain
version of the) Hamiltonian of the supersymmetric KY model in the subspace~$n^0=0$ is given by
\begin{equation}
  \label{HKY2}
  \hat H|_{\cN_0=0} = \frac1{2N^2}\sum_{i<j}\frac{1+P_{ij}}{\sin^2(\pi(i-j)/N)}
  -\frac h2(\cN_1-\cN_2)-\mu N\,,
\end{equation}
where~$P_{ij}$ is the ordinary permutation operator. This should be compared with the Hamiltonian
of the (antiferromagnetic) $\su(2)$ HS chain in an external magnetic field from Ref.~\cite{EFG12}
(with~$K=-1$ and $2B=h$), namely
\begin{equation}
  \label{H2mag}
  \hat H^{(2)}= \frac1{2N^2}\sum_{i<j}\frac{P_{ij}-1}{\sin^2(\pi(i-j)/N)}
  -\frac h2(\cN_1-\cN_2)\,.
\end{equation}
From Eq.~\eqref{sinsum} it follows that in the thermodynamic limit
$\hat H|_{\cN_0=0} = \hat H^{(2)}-\mu_cN$. It should therefore be expected that in the vertical
band~$\cB_1$, and at sufficiently low temperatures,
\[
  \fl
  f(T)\simeq f^{(2)}(T)-\mu_c=-\mu_c-2T\int_0^{1/2}\log\Bigl[\cosh\bigl(\tfrac{\be h}2\bigr)
  +\sqrt{\sinh^2\bigl(\tfrac{\be h}2\bigr)+\e^{\be\vep(x)}}\,\Bigr]\diff x\equiv g(T),
\]
where we have used the exact formula for~$f^{(2)}$ from Ref.~\cite{EFG12}. The asymptotic series
of~$g$ around $T=0$ can be obtained following the above procedure. More precisely, we first
write
%{\interdisplaylinepenalty=10000
\begin{eqnarray*}
  \fl
  g(T)=g(0)
  &-2T\int_0^{x_0(h)}\log\Bigl[\tfrac12(1+\e^{-\be h})+
    \sqrt{\tfrac14(1-\e^{-\be h})^2+\e^{-\be(h-\vep(x))}}\,\Bigr]\diff x\\
  \fl
  &-2T\int_{x_0(h)}^{1/2}\log\Bigl[\tfrac12(1+\e^{-\be h})\e^{-\frac\be2(\vep(x)-h)}+
    \sqrt{1+\tfrac14(1-\e^{-\be h})^2\e^{-\be(\vep(x)-h)}}\,\Bigr]\diff x\,,
\end{eqnarray*}
%}
where
\[
  g(0)=-\mu_c-hx_0(h)-\int_{x_0(h)}^{1/2}\vep(x)\,\diff x=f(0)
\]
(cf.~Eq.~\eqref{f0B1}). Discarding the exponentially small term~$\e^{-\be h}$, and comparing with
Eqs.~\eqref{I1def2}-\eqref{I2def2}, we thus see that
\[
  g(T)\sim f(0)-2T(I_1+I_2)\,.
\]
From Eqs.~\eqref{f3} and~\eqref{I1def}-\eqref{I2def} we conclude that~$g$ has the same asymptotic
series as~$f$ in the vertical band~$\cB_1$, as stated.

\section{Low-temperature asymptotic expansions of densities and susceptibilities}\label{sec.LTE}

Differentiating the asymptotic series for the free energy per site obtained in the previous
section with respect to the parameters $h$ and $\mu$, we shall next to derive analogous series for
the magnetization per site, the charge density and their corresponding susceptibilities. Since the
asymptotic series of all these quantities are trivially equal to their zero temperature values on
the wedges $\cW_0$ and $\cW_1$, we shall concentrate in what follows on the remaining
regions~$\cB_0$, $\cB_1$ and~$\cT$.

\subsection{$\cB_1\cup\cT$}
In the region $\cB_1\cup \cT$, by
Eq.~\eqref{psi} the asymptotic series for the magnetization is given by
\[
m_s\sim-\frac{\partial f(0)}{\partial h}-\frac{\partial\psi(T,h)}{\partial h}.
\]
From Eqs.~\eqref{f0B1} and~\eqref{f0T} it follows that
\[
-\frac{\partial f(0)}{\partial h}=x_0(h)\,,
\]
which coincides with the value of $m_s(0)$ computed in Section~\ref{sec.T0}. Using
Eq.~\eqref{psiTh} we thus obtain
\begin{equation}
  \label{msasyB1T}
  m_s\sim x_0(h)+\sum_{l=0}^\infty \frac{2^{l+2}(2l+1)!!}{l!\,{(1-4h)}^{l+\frac32}}\,
  \mathcal I_l\,T^{l+2},
\end{equation}
where $\mathcal I_l$ is given by~Eq.~\eqref{cIl}. The first few terms in this series are thus
\[
  \fl
  m_s=x_0(h)+\frac{2\pi^2T^2}{3(1-4h)^{3/2}}+\frac{24\ze(3)T^3}{(1-4h)^{5/2}}
  +\frac{12\pi^4T^4}{(1-4h)^{7/2}}+\Or(T^5),
\]
Differentiating Eq.~\eqref{msasyB1T} with respect to $h$ we obtain the corresponding asymptotic
series for the magnetic susceptibility $\chi_s$: 
\begin{equation}
  \chi_s\sim (1-4h)^{-1/2}+\sum_{l=0}^\infty\frac{2^{l+3}(2l+3)!!}{l!\,{(1-4h)}^{l+\frac52}}\,
  \mathcal I_l\,T^{l+2}.
  \label{chisasyT}
\end{equation}
To the best of our knowledge, for $(h,\mu)$ lying on the vertical band~$\cB_1$ the asymptotic
expansions~\eqref{msasyB1T}-\eqref{chisasyT} are new.

Consider next the charge density $n_c$ and its susceptibility $\chi_c$. To begin with, it is
obvious that $n_c\sim 1$ and $\chi_c\sim0$ in the vertical band $\cB_1$, since in this region
$f(0)+\mu$ and the asymptotic series for $f(T)-f(0)$ depend only on $h$ (cf.~Eqs.~\eqref{f0B1}
and~\eqref{fasyB1}). On the other hand, in the triangle $\cT$ the asymptotic series of $n_c$ and
$\chi_c$ follow immediately from those of $m_s$ and $\chi_s$ taking into account Eq.~\eqref{psi},
namely
\begin{eqnarray}
  &n_c\sim 2x_0(2\mu)+\sum_{l=0}^\infty \frac{2^{l+3}(2l+1)!!}{l!\,{(1-8\mu)}^{l+\frac32}}\,\mathcal
    I_l\,(-T)^{l+2},\label{ncasyT}\\
  &\chi_c\sim4(1-8\mu)^{-1/2}+\sum_{l=0}^\infty\frac{2^{l+5}(2l+3)!!}{l!\,{(1-8\mu)}^{l+\frac52}}\,
    \mathcal I_l\,(-T)^{l+2}, \quad (h,\mu)\in\cT.\label{chicasyT}
\end{eqnarray}
In fact, Eqs.~\eqref{chisasyT} and~\eqref{chicasyT} agree to all orders in~$T$ with the asymptotic
series which can be obtained from the low-temperature approximations (up to exponentially small
terms) to~$m_s$ and~$n_c$ in Ref.~\cite{KK95}. To see this, note first that in our notation the
latter approximations read
\[
  \fl 1-n_c\sim2\int_{-\infty}^\infty\frac{\sqrt{1-8\mu+4Tx}}{(1+4\e^x)^{3/2}}\,\e^x\diff x\,,\qquad
  1-2m_s\sim2\int_{-\infty}^\infty\frac{\sqrt{1-4h-4Tx}}{(1+4\e^x)^{3/2}}\,\e^x\diff x\,.
\]
Expanding the square root in either integral in powers of~$T$ and using the alternative
expression~\eqref{Ilalt} for the integrals~$\cI_l$ we readily obtain Eqs.~\eqref{msasyB1T}
and~\eqref{ncasyT}, from which the corresponding asymptotic series for~$\chi_s$ and~$\chi_c$
follow by term-by-term differentiation.

From the previous formulas it follows that in the triangle~$\cT$ the magnetization~$m_s$ and its
susceptibility~$\chi_s$ depend only on~$h$, while $n_c$ and~$\chi_c$ depend only on~$\mu$, up to
terms~$\Or(T^k\e^{-c\be})$ (with $c>0$). This is an indication that spin-charge separation is
valid at low temperatures up to terms exponentially small in~$\be$, as we shall more explicitly
show in what follows.
\begin{figure}[t]
  \includegraphics[width=.49\textwidth]{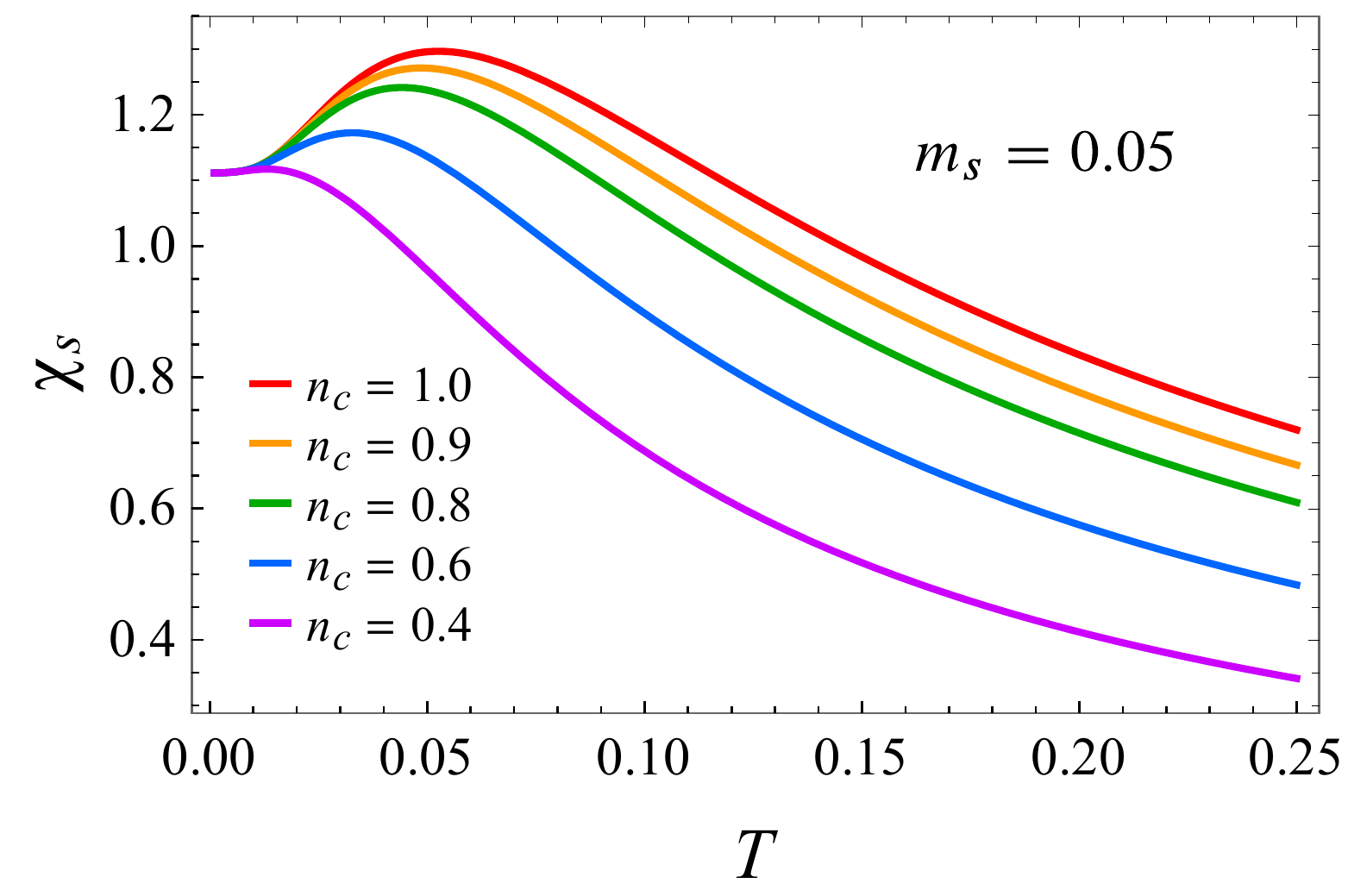}\hfill
  \includegraphics[width=.49\textwidth]{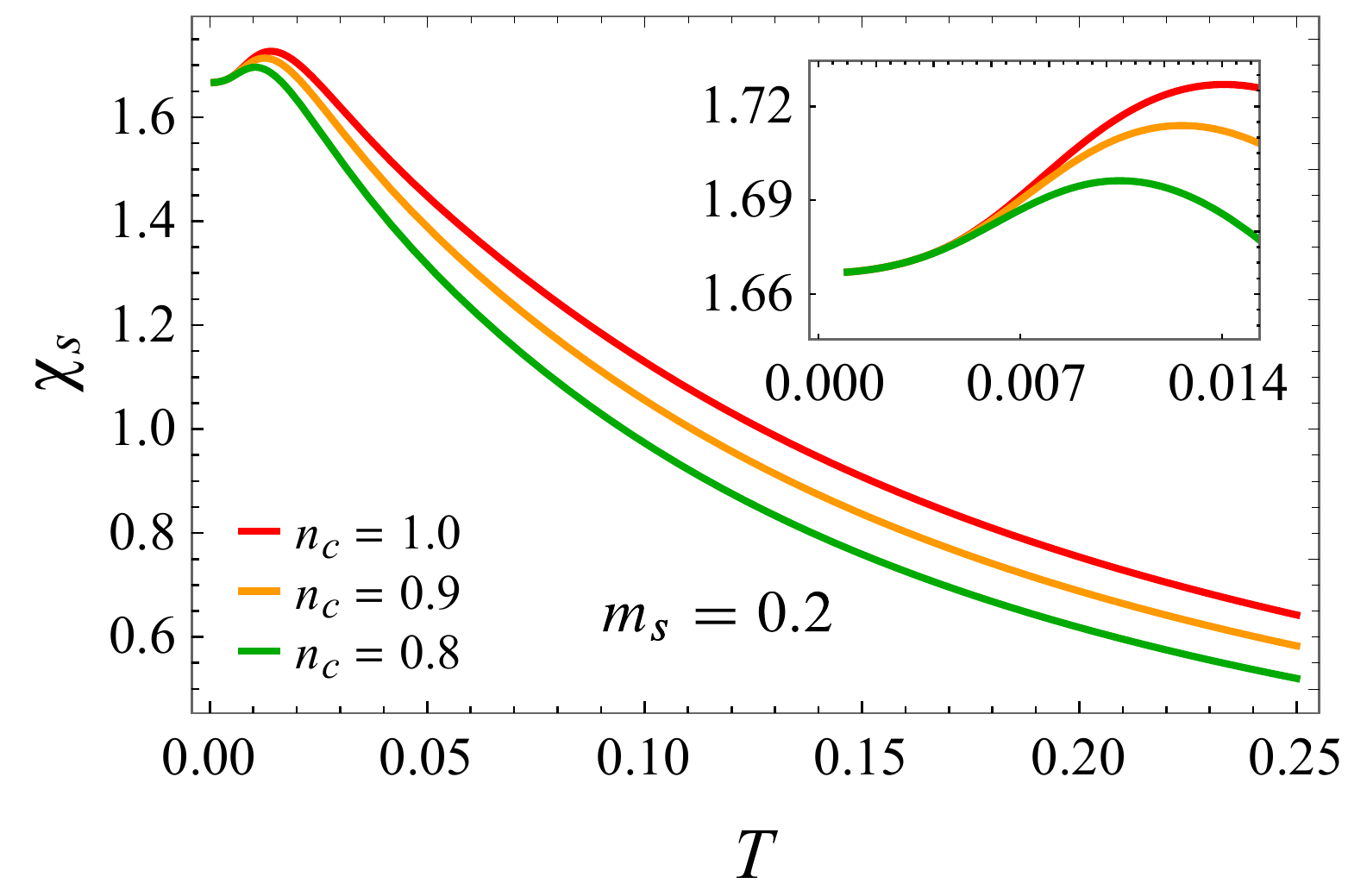}
  \caption{Magnetic susceptibility per site~$\chi_s$ vs.~temperature~$T$ at constant charge
    density~$n_c$ and magnetization $m_s=0.05$ (left) and~$m_s=0.2$ (right). Inset: detail of the
    plot in the smaller range~$0\le T\le0.015$. In all plots, $T$ and $1/\chi_s$ are measured in
    units of~$K=2\pi^2 t$.}
  \label{fig.chis}
\end{figure}

\begin{figure}[b]
  \includegraphics[width=.49\textwidth]{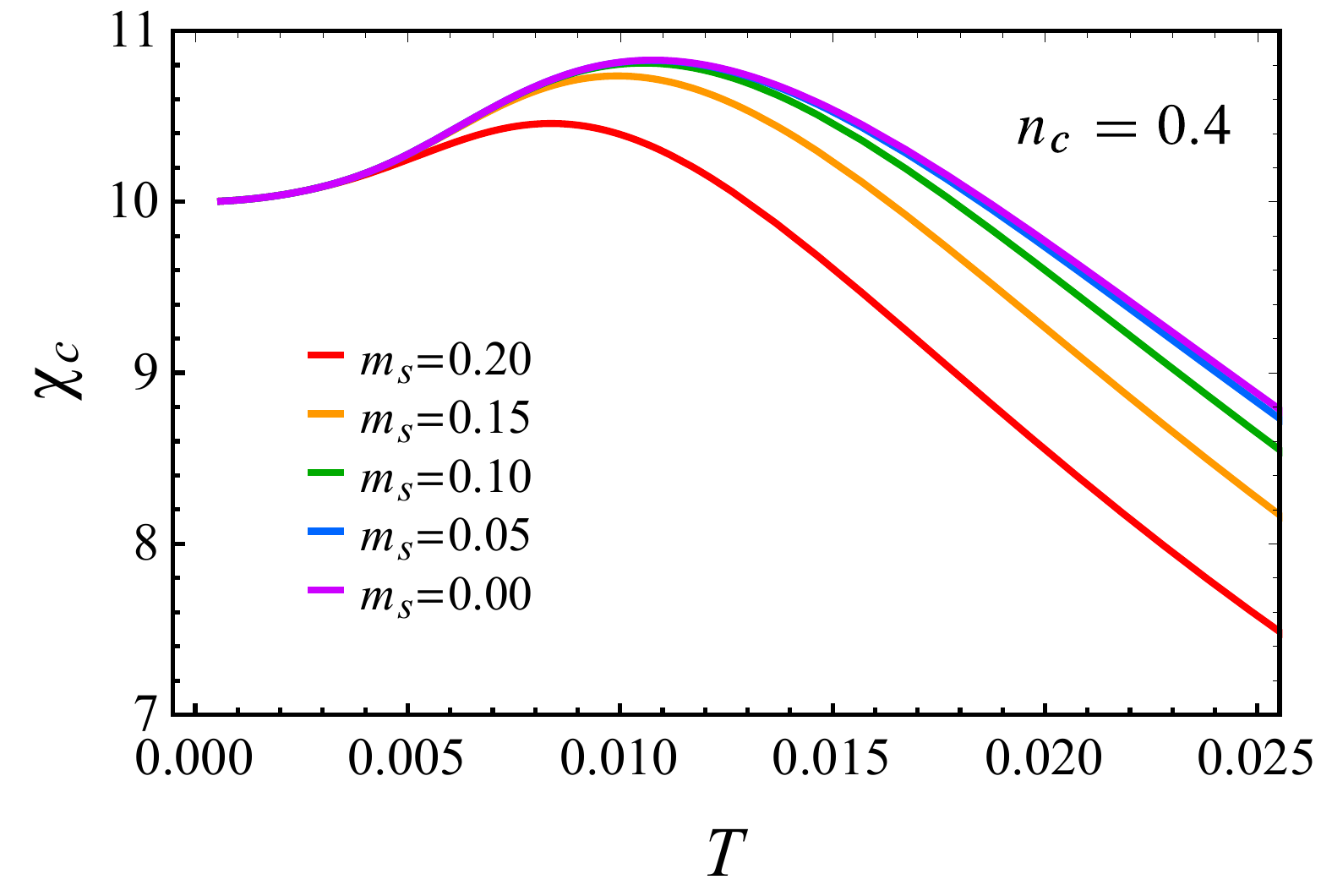}\hfill
  \includegraphics[width=.49\textwidth]{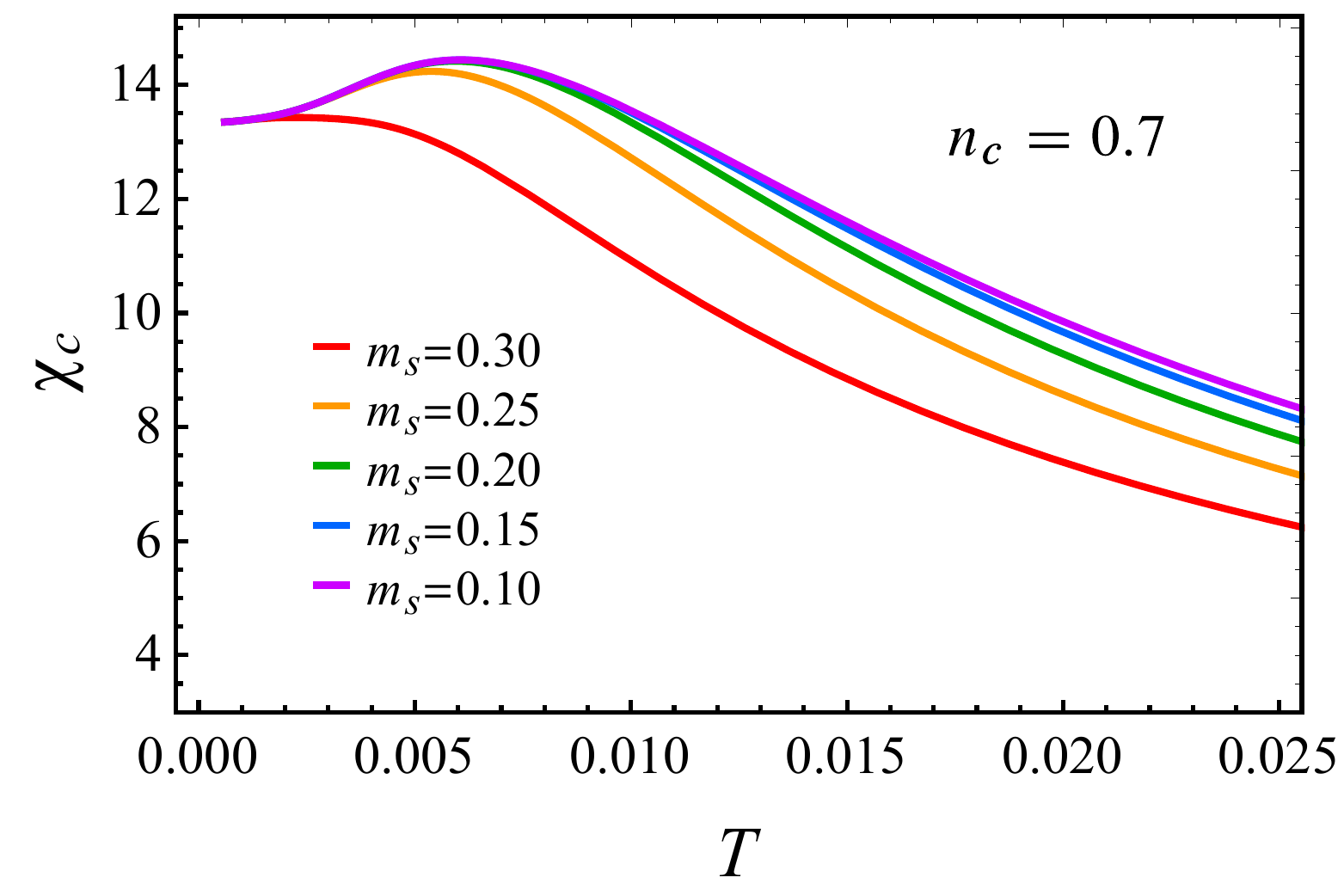}
  \caption{Charge susceptibility per site vs.~temperature~$T$ at constant magnetization
    density~$m_s$ and charge density $n_c=0.4$ (left) and~$0.7$ (right). In all plots, $T$
    and~$1/\chi_c$ are measured in units of $K=2\pi^2t$.}
  \label{fig.chic}
\end{figure}

Indeed, inverting the asymptotic expansion~\eqref{msasyB1T} of~$m_s$ to obtain an expansion
of~$1-2m_s(0)\equiv(1-4h)^{1/2}$ to a given order in $T$ and substituting into
Eq.~\eqref{chisasyT} we can derive a corresponding asymptotic expansion of~$\chi_s$ in terms
of~$m_s$. For instance, up to third order in $T$ we have
\begin{equation}
  \fl
  \chi_s=\frac1{1-2m_s}\Bigg[1+\frac{8\pi^2T^2}{3(1-2m_s)^4}+
  \frac{192\ze(3)T^3}{(1-2m_s)^6}\Bigg]+\Or(T^4).
  \label{chisX}
\end{equation}
Comparing the asymptotic series~\eqref{ncasyT}-\eqref{chicasyT} for $n_c$ and $\chi_c$ with the
corresponding series~\eqref{msasyB1T}-\eqref{chisasyT} for $m_s$ and $\chi_s$ we conclude that the
asymptotic series of $\chi_c$ in the triangle~$\cT$ is given by $\chi_c\sim 4X(\tfrac{n_c}2,-T)$,
where $X(m_s,T)$ is the asymptotic series for $\chi_s$ in terms of $m_s$ and $T$. From
Eq.~\eqref{chisX} we thus obtain
\begin{equation}
  \fl
  \chi_c=\frac4{1-n_c}\Bigg[1+\frac{8\pi^2T^2}{3(1-n_c)^4}-
  \frac{192\ze(3)T^3}{(1-n_c)^6}\Bigg]+\Or(T^4),\qquad
  (h,\mu)\in\cT.\label{chicX}
\end{equation}
From the asymptotic expansions~\eqref{chisX}-\eqref{chicX} it is clear that strong spin-charge
separation holds at sufficiently low temperatures if we discard exponentially small terms.
Previously, this result had been numerically checked only for~$h=0$ (i.e., $m_s=0$) and within the
triangle~$\cT$~\cite{KK95,KK96}. With the help of the exact formulas~\eqref{chis}-\eqref{chic} for
the susceptibilities derived in Section~\ref{sec.F}, we have been able to numerically verify the
strong spin-charge separation for non-zero magnetic fields and several charge densities
(see~Figs.~\ref{fig.chis}-\ref{fig.chic}). As is customary, we have taken $(m_s,n_c)$ instead of
$(h,\mu)$ as independent variables, which requires solving for the latter variables in terms of
the former by means of Eqs.~\eqref{ms}-\eqref{nf}. For this to be possible, the mapping
$(h,\mu)\mapsto(m_s,n_c)$ should be invertible in the temperature range considered. At zero
temperature, this is the case provided that $(h,\mu)$ lies in the triangle~$\cT$, which is mapped
to the triangle~$0<2m_s<n_c<1$. For this reason, in the previous plots we have taken
magnetizations not greater than $n_c/2$.

Again, when~$(h,\mu)\in\cB_1$ the asymptotic expansion~\eqref{chisX} appears to be new, while in
the triangle~$\cT$ Eqs.~\eqref{chisX}-\eqref{chicX} coincide with those derived in
Ref.~\cite{KK95} to order~$T^2$. (To verify this assertion one should first replace the
dimensionless quantities $T$, $\chi_s$ and $\chi_c$ in the latter equations by their true values
$T/K$, $K\chi_s$ and $K\chi_c$, with $K=2\pi^2t$, and take into account that in the latter
reference~$t$ has been set to~$1$.) Note also that the asymptotic expansions of~$\chi_s$ and
$\chi_c$ do \emph{not} have the same functional form, due to the different sign of the
coefficients of the odd powers of~$T$. This fact had not been previously noted, since it can only
be detected at order~$T^3$ or higher.

\subsection{$\cB_0$}
Consider, finally, the oblique band~$\cB_0$. The asymptotic series for $m_s$ and $\chi_s$ are
easily obtained differentiating Eqs.~\eqref{f0B0} and~\eqref{fasyB0}, i.e.,
\begin{eqnarray}
  \fl
  &m_s\sim x_0(\tfrac
       h{2}+\mu)+2\sum_{l=0}^\infty\frac{(2^{2l+1}-1)(4l+1)!!}{\big[1-2(h+2\mu)\big]^{2l+\frac32}}\,
       \ze(2l+2)\,T^{2l+2},\label{msasyB0}\\
  \fl
  &\chi_s\sim\frac1{2}\,\big[1-2(h+2\mu)\big]^{-1/2}+2\sum_{l=0}^\infty\frac{(2^{2l+1}-1)(4l+3)!!}%
           {\big[1-2(h+2\mu)\big]^{2l+\frac52}}\,\ze(2l+2)\,T^{2l+2}.
  \label{chisasyB0}
\end{eqnarray}
Proceeding as above, it is straightforward to derive from the previous series an asymptotic
expansion of~$\chi_s$ in terms of the variables $(m_s,T)$ in the oblique band~$B_0$ to any desired
order. The first few terms in this expansion are
\begin{equation}
  \fl
  \chi_s\sim\frac1{2(1-2m_s)}\bigg[1+\frac{4\pi^2T^2}{3(1-2m_s)^4}
    +\frac{28\pi^4T^4}{9(1-2m_s)^8}
    +\frac{3352\pi^6T^6}{27(1-2m_s)^{12}}
    +\Or(T^8)\bigg]\,.
    \label{chisms}
\end{equation}
From Eqs.~\eqref{f0B0} and~\eqref{fasyB0} it follows that the asymptotic series of the free energy
is a function of $h+2\mu$. Thus the asymptotic series of $n_c$ and $\chi_c$ are proportional to
those of $m_s$ and $\chi_s$, namely,
\[
  n_c\sim 2m_s,\qquad \chi_c\sim4\chi_s\,.
\]
Thus in the band~$\cB_0$ the magnetic and the charge quantities are proportional, up to
exponentially small terms in~$\be$. In particular, from Eq.~\eqref{chisms} we obtain the following
asymptotic expansion of~$\chi_c$ in terms of~$1-n_c$ in the band~$\cB_0$:
\begin{equation}
  \fl
  \chi_c\sim\frac2{1-n_c}\bigg[1+\frac{4\pi^2T^2}{3(1-n_c)^4}
    +\frac{28\pi^4T^4}{9(1-n_c)^8}
    +\frac{3352\pi^6T^6}{27(1-n_c)^{12}}
    +\Or(T^8)\bigg]\,.
    \label{chicnc}
\end{equation}

\section{Conclusions}\label{sec.conc}

In this paper we analyze the thermodynamics of the supersymmetric $\su(m)$ $t$-$J$ model with
long-range interactions through a novel approach based on the transfer matrix method. This method
exploits the equivalence of the latter model to a modification of the $\su(1|m)$ supersymmetric
Haldane--Shastry spin chain, whose spectrum coincides with that of an inhomogeneous vertex model
with a simple dispersion function. The energy function of this vertex model is related to suitable
representations of the Yangian associated to supersymmetric Young tableaux and their corresponding
Haldane motifs. This makes it possible to express the partition function by means of an
appropriate site-dependent transfer matrix, which in the thermodynamic limit yields a simple
closed-form expression for the free energy per site in terms of the largest eigenvalue (in
modulus) of the latter matrix. One of the main advantages of our method is the fact that it can be
applied to a wide range of models with (broken or unbroken) Yangian symmetry and arbitrary
dispersion relations, including the supersymmetric Polychronakos--Frahm and Frahm--Inozemtsev spin
chains. In the $\su(1|m)$ case analyzed in the paper, we explicitly show that the free energy per
site of all of these models can be expressed in terms of a function of one variable obeying an
algebraic equation which generalizes the one derived by Kato and Kuramoto for multi-component
boson-fermion systems~\cite{KK96,KK09}. We conjecture that this is still the case for more
general~$\su(n|m)$ models with $n>1$.

In the spin $1/2$ case, we apply the explicit expression for the free energy to analyze in detail
the thermodynamic and criticality properties of the model. To this end, we first determine all the
ground state phases by computing the zero-temperature values of the magnetization and charge
densities for arbitrary values of the magnetic field strength and the charge chemical potential.
In particular, we show that the magnetic and charge susceptibilities present hitherto unnoticed
jump discontinuities along the common boundary of the~$\su(1|2)$ and $\su(1|1)$ phases. We then
derive the complete asymptotic series of the free energy per site, showing that it takes different
forms on each of the ground state phases. From the lowest-order term in the asymptotic series we
determine the regions in parameter space in which the model is described at low energies by an
effective CFT, and compute its corresponding central charge. Our results confirm that in the
$\su(1|2)$ phase the model is described by a CFT with conformal charge $c=1$ in both the spin and
the charge sectors. However, in the $\su(2)$ and $\su(1|1)$ phases we find that the model is
equivalent to a single CFT with $c=1$. We also analyze in detail the critical behavior on the
boundary between zero-temperature phases, finding that the system can be critical, gapless but not
critical or even critical in the spin sector but not in the charge one. Using the asymptotic
series for the free energy, we also derive the complete asymptotic series of the magnetization and
charge densities and their corresponding susceptibilities. We numerically verify the strong
spin-charge separation characteristic of the model for different (nonzero) values of the
magnetization and the charge density, and show that it persists at all orders in the asymptotic
expansion. This can be regarded as an analytic confirmation of spin-charge separation in a
sufficiently small range of temperatures near $T=0$, where the asymptotic expansions provide an
excellent approximation for the thermodynamic functions.

Although in this paper we have concentrated on the $\su(2)$ case, it would be of interest to apply
the transfer matrix method to investigate the ground-state phases, thermodynamics and criticality
properties of the general $\su(m)$ KY model with $m>2$. In fact, as explained above, this method
could in principle be extended to more general models with partial or total Yangian symmetry
provided that their spectrum coincides with that of the inhomogeneous vertex
model~\eqref{Evertex}-\eqref{demn} for a suitable dispersion relation~$\cE_N$.

\section*{Acknowledgments}
This work was partially supported by Spain's MINECO grant~FIS2015-63966-P and the Universidad
Complutense de Madrid through a G/6400100/3000 grant. JAC would also like to acknowledge the
financial support of the Universidad Complutense de Madrid through a 2015 predoctoral scholarship.

\appendix

\section{Derivation of Eqs.~\eqref{I1def}-\eqref{I2def}}\label{app.A}
In this appendix we shall establish the validity of the estimates~\eqref{I1def}-\eqref{I2def} for
the integrals appearing in Eq.~\eqref{f3}. To begin with, the difference between the first of
these integrals and the integral~$I_1$ in Eq.~\eqref{I1def} is given by
\begin{equation}
  \fl
  \De_1\equiv\int_0^{x_0(h)}\kern-.5em\log\left[\frac{\tb+\sqrt{\vphantom{\tfrac14}
        \tb^2+\e^{-\be(h-\vep)}-\e^{-\be h}}}%
    {\frac12+\sqrt{\frac14+\e^{-\be(h-\vep)}}}\,\right]\diff x
  =\int_0^{x_0(h)}\kern-1.2em\log\bigl(1+\phi_1(x)\bigr)\diff x\,,
         \label{De1}
\end{equation}
where
\begin{equation}\label{phi}
  \phi_1(x)\equiv\frac{\frac12(\e^{-\be(\mu+\frac h2-\vep)}+\e^{-\be h})+\De R_1}{\frac12
    +\sqrt{\frac14+\e^{-\be(h-\vep)}}}
\end{equation}
and
\[
  \fl \De R_1\equiv \sqrt{\tb^2+\e^{-\be(h-\vep)}-\e^{-\be h}}-\sqrt{\tfrac14+\e^{-\be(h-\vep)}}
  =\frac{\tb^2-\frac14-\e^{-\be h}}{\sqrt{\vphantom{\tfrac14}\tb^2+\e^{-\be(h-\vep)}-\e^{-\be
        h}}+\sqrt{\tfrac14+\e^{-\be(h-\vep)}}}\,.
\]
Since~$\tb>1/2$ the denominator in~$\De R_1$ is~$\ge1$, and hence
\begin{eqnarray*}
  \fl
  |\De R_1|&\le\e^{-\be h}+\tb^2-\tfrac14=\e^{-\be h}
             +\frac14\Big(\e^{-\be(\mu+\frac h2-\vep)}+\e^{-\be h}\Big)\Big(2+\e^{-\be(\mu+\frac
             h2-\vep)}+\e^{-\be h}\Big)\\
  \fl
           &\le \e^{-\be(\mu-\frac h2)}+2\e^{-\be h},
\end{eqnarray*}
where we have used the inequality~$\vep(x)\le h$ valid in the interval~$[0,x_0(h)]$. From
Eq.~\eqref{phi} we thus obtain
\[
  |\phi_1(x)|\le \tfrac32\,\e^{-\be(\mu-\frac h2)}+\tfrac52\,\e^{-\be
    h}=\Or\bigl(\e^{-\be\min(\mu-\frac h2,h)}\bigr) \,,
\]
which in particular shows that when~$(h,\mu)\in \cB_1$ the function~$\phi_1(x)$ tends to zero as
$T\to0$ uniformly in~$x\in[0,x_0(h)]$. Since $|\log(1+\phi_1(x))|=\Or(\phi_1(x))$
when~$\phi_1(x)\to0$, from Eq.~\eqref{De1} it immediately follows that
\[
  \De_1=\Or\bigl(\e^{-\be\min(\mu-\frac h2,h)}\bigr).
\]

Similarly, the difference between the LHS of Eq.~\eqref{I2def} and the
integral~\eqref{I2def} can be written as
\begin{equation}\label{De2}
  \De_2=\int_{x_0(h)}^{1/2}\kern-1em\log\bigl(1+\phi_2(x)\bigr)\diff x\,,
\end{equation}
where
\begin{eqnarray*}
  \fl
  \phi_2(x)&\equiv\frac{\frac12(\e^{-\frac\be2(\vep+h)}+\e^{-\be(\mu-\frac{\vep}2)}) +\De
    R_2}{\frac12\e^{-\frac\be2(\vep-h)} +\sqrt{1+\frac14\e^{-\be(\vep-h)}}}\\
  \fl \De R_2&\equiv \sqrt{\hb^2+1-\e^{-\be \vep}}-\sqrt{1+\tfrac14\e^{-\be(\vep-h)}}
  =\frac{\hb^2-\frac14\e^{-\be(\vep-h)}-\e^{-\be \vep}}{\sqrt{\vphantom{\tfrac14} \hb^2+1-\e^{-\be
        \vep}}+\sqrt{1+\tfrac14\e^{-\be(\vep-h)}}}\,.
\end{eqnarray*}
Proceeding as before, and taking into account that in the interval~$[x_0(h),1/2]$ we have
\[
  \vep(x)\ge h\,,\qquad \mu-\frac{\vep(x)}2\ge \mu-\frac 18\,,
\]
after a straightforward calculation we obtain the estimate
\[
  \Big|\,\hb^2-\tfrac14\e^{-\be(\vep-h)}-\e^{-\be \vep}\Big|\le
  \e^{-\frac\be2(\vep+h)}+\e^{-\be(\mu-\frac\vep2)}+\e^{-\be\vep}\le 2\e^{-\be
    h}+\e^{-\be(\mu-\frac 18)}\,.
\]
From the definition of~$\phi_2(x)$ it immediately follows that
\[
  |\phi_2(x)|\le\tfrac32\,\e^{-\be(\mu-\frac 18)}+\tfrac52\,\e^{-\be h}=
  \Or\bigl(\e^{-\be\min(\mu-\frac18,h)}\bigr),
\]
which easily yields~\eqref{I2def} on account of Eq.~\eqref{De2}.

\section{Asymptotic series for the integral $I_1$}\label{app.B}

In this appendix we derive the asymptotic series~\eqref{I1asymp} for the integral $I_1$. Calling
for simplicity~$a_l\equiv a_l(h)$ and setting
\[
  g(y)\equiv\log\Bigl[\tfrac12+\sqrt{\tfrac14+\e^{-y}}\,\Bigr],\qquad
  \phi(z)\equiv \sum_{l=0}^\infty(-1)^la_lz^l,
\]
we need to show that
\[
  \int_0^{\be h}g(y)\phi(Ty)\diff y\sim\sum_{l=0}^\infty(-1)^la_l\,T^l\!\!\int_0^\infty
  y^lg(y)\diff y\,.
\]
Note first of all that the power series $\phi(z)$ converges for $|z|<1/4$. Since $h<1/4$ when
$(h,\mu)\in \cB_1$, it follows that $Ty$ lies inside the convergence disc of $\phi(z)$ for fixed
$h$ and all $y\in[0,\be h]$. We must check that for all $n\in\NN$
\[
  \sum_{l=0}^n(-1)^la_l\,T^l\!\!\int_0^\infty y^lg(y)\diff y -\int_0^{\be h}g(y)\phi(Ty)\diff
  y=\ord(T^n)\,,
\]
i.e.,
\begin{equation}\label{oTn}
  \fl
  \sum_{l=0}^n(-1)^la_l\,T^l\int_{\be h}^\infty y^lg(y)\diff y
  -\int_0^{\be h} \diff y\,g(y)\sum_{l=n+1}^\infty(-1)^la_l\,(Ty)^l=\ord(T^n).
\end{equation}
Since
\[
  0\le\int_{\be h}^\infty y^lg(y)\diff y\le \int_{\be h}^\infty y^l\e^{-y}\diff
  y=\Or(\be^l\e^{-\be h}),
\]
the first sum in Eq.~\eqref{oTn} is $\Or(\e^{-\be h})$. As to the second term, note that
\[
  \sum_{l=n+1}^\infty(-1)^la_l\,(Ty)^l=(Ty)^{n+1}\,\widetilde\phi(Ty),
\]
where $\widetilde\phi(z)$ is a convergent power series and hence analytic for $|z|<1/4$. Since
$Ty\in[0,h]\subset[0,1/4)$ when $y\in[0,\be h]$, it follows that $ |\widetilde\phi(Ty)|<M(h) $
independently of $\be$. Hence
\begin{eqnarray*}
  \fl
  \bigg|\int_0^{\be h} \diff y\,g(y)
  \sum_{l=n+1}^\infty(-1)^la_l\,(Ty)^l\bigg|
  &\le M(h)\,T^{n+1}\!\!\int_0^{\be h} y^{n+1}g(y)\diff y\\
  \fl
  &\le \bigg(M(h)\int_0^{\infty} y^{n+1}e^{-y}\diff y\bigg)\,T^{n+1}=(n+1)!M(h)\,T^{n+1},
\end{eqnarray*}
so that both terms in the LHS of Eq.~\eqref{oTn} are indeed $\ord(T^n).$

\section{Alternative expression for the integrals~$\cI_l$}\label{app.ints}
In this appendix we will derive the alternative expressions~\eqref{Ilalt} for the
integrals~$\cI_l$ appearing in the asymptotic series for the free energy in the triangle~$\cT$ and
the vertical band~$\cB_1$ (cf.~Eq.~\eqref{cIl}). To begin with, consider the integral
\[
  \cI_{l,1}\equiv\int_0^{\infty}y^l\Big\{\log\Bigl[\tfrac12\,\e^{-y/2}
  +\sqrt{1+\tfrac14\,\e^{-y}}\,
  \Bigr]\Big\}\diff y=
  \int_0^\infty y^l\arcsinh\bigl(\tfrac12\,\e^{-y/2}\bigr)\,\diff y\,.
\]
Writing $\arcsinh\bigl(\frac12\,\e^{-y/2}\bigr)$ as the integral of its derivative, namely
\[
\arcsinh\bigl(\tfrac12\,\e^{-y/2}\bigr)=\frac12\int_y^\infty\frac{\diff x}{\sqrt{1+4\e^x}}\,,
\]
we can express~$\cI_{l,1}$ as a double integral as
\[
  \cI_{l,1}=\frac12\int_0^{\mathrlap{\infty}}\diff y\int_y^{\mathrlap{\infty}}\diff
  x\,\frac{y^l}{\sqrt{1+4\e^x}}\,.
\]
Reversing the order of integration we obtain
\begin{equation}
  \label{Il1}
  \cI_{l,1}=\frac12\int_0^{\mathrlap{\infty}}\frac{\diff x}{\sqrt{1+4\e^x}}\int_0^{\mathrlap{x}}\diff
  y\,y^l=\frac1{2(l+1)}\int_0^{\mathrlap{\infty}}\frac{x^{l+1}}{\sqrt{1+4\e^x}}\,\diff x\,.
\end{equation}
Consider next the second integral in Eq.~\eqref{cIl}, namely
\[
  \cI_{l,2}=\int_0^{\infty}y^l\Big\{\log\Bigl[\tfrac12 +\sqrt{\tfrac14+\e^{-y}}\, \Bigr]\Big\}\diff y
  =\int_0^{\infty}y^l\Big[-\tfrac y2+\arcsinh\bigl(\tfrac12\,\e^{y/2}\bigr)\Big]\diff y\,.
\]
Proceeding as before we write
\[
  -\frac y2+\arcsinh\bigl(\tfrac12\,\e^{y/2}\bigr)=
  \frac12\int_y^\infty\left(1-\frac{1}{\sqrt{1+4\e^{-x}}}\right)\diff x
\]
and therefore
\begin{eqnarray}
  \fl
  \cI_{l,2}
  &=\frac12\int_0^{\mathrlap{\infty}}\diff y\int_y^{\mathrlap{\infty}}\diff
    x\,y^l\left(1-\frac{1}{\sqrt{1+4\e^{-x}}}\right)
    =\frac12 \int_0^{\mathrlap{\infty}}\diff x\left(1-\frac{1}{\sqrt{1+4\e^{-x}}}\right)\int_0^{x}\diff
    y\,y^l\nonumber\\
  \fl
  &=\frac{(-1)^l}{2(l+1)}\int_{-\mathrlap{\infty}}^0x^{l+1}\left(\frac1{\sqrt{1+4\e^x}}-1\right)
    \diff x\,.
    \label{Il2}
\end{eqnarray}
Combining Eqs.~\eqref{Il1} and~\eqref{Il2} we obtain the first equality in Eq.~\eqref{Ilalt}. The
second equality in the latter equation easily follows integrating by parts in Eqs.~\eqref{Il1}
and~\eqref{Il2}.

\section*{References}

% \bibliographystyle{jsm-te} 
% \bibliography{cmprefs}

\end{document}